%% file: main.tex
\documentclass[sigconf, nonacm]{acmart}

\newcommand\vldbdoi{XX.XX/XXX.XX}
\newcommand\vldbpages{XXX-XXX}
\newcommand\vldbvolume{14}
\newcommand\vldbissue{1}
\newcommand\vldbyear{2020}
\newcommand\vldbauthors{\authors}
\newcommand\vldbtitle{\shorttitle} 
\newcommand\vldbavailabilityurl{URL_TO_YOUR_ARTIFACTS}
\newcommand\vldbpagestyle{plain} 

\input{secs/commands}
\input{secs/notations}

\begin{document}

\title{QSPN: Query-aware Sum-Product Network for Comprehensive Advantage Cardinality Estimation}
\title{A Unified Model for Cardinality Estimation by Learning from Data and Queries via Sum-Product Networks}

\settopmatter{authorsperrow=4}
\author{Jiawei Liu}
\affiliation{%
	\institution{Renmin University of China}
}
\email{jiaweiliu@ruc.edu.cn}

\author{Ju Fan}
\affiliation{%
	\institution{Renmin University of China}
}
\email{fanj@ruc.edu.cn}

\author{Tongyu Liu}
\affiliation{%
	\institution{Renmin University of China}
}
\email{ltyzzz@ruc.edu.cn}

\author{Kai Zeng}
\affiliation{%
	\institution{Huawei Technologies}
}
\email{kai.zeng@huawei.com}

\author{Jiannan Wang}
\affiliation{%
	\institution{Huawei Technologies \& Simon Fraser University}
}
\email{jnwang@sfu.ca}

\author{Quehuan Liu}
\affiliation{%
	\institution{Renmin University of China}
}
\email{liuquehuan@ruc.edu.cn}

\author{Tao Ye}
\affiliation{%
	\institution{Huawei Technologies}
}
\email{yetao1@huawei.com}

\author{Nan Tang}
\affiliation{%
	\institution{HKUST(GZ)}
}
\email{nantang@hkust-gz.edu.cn}

\input{secs/abstract}

\maketitle

\pagestyle{\vldbpagestyle}
\begingroup\small\noindent\raggedright\textbf{PVLDB Reference Format:}\\
\vldbauthors. \vldbtitle. PVLDB, \vldbvolume(\vldbissue): \vldbpages, \vldbyear.\\
\href{https://doi.org/\vldbdoi}{doi:\vldbdoi}
\endgroup
\begingroup
\renewcommand\thefootnote{}\footnote{\noindent
	This work is licensed under the Creative Commons BY-NC-ND 4.0 International License. Visit \url{https://creativecommons.org/licenses/by-nc-nd/4.0/} to view a copy of this license. For any use beyond those covered by this license, obtain permission by emailing \href{mailto:info@vldb.org}{info@vldb.org}. Copyright is held by the owner/author(s). Publication rights licensed to the VLDB Endowment. \\
	\raggedright Proceedings of the VLDB Endowment, Vol. \vldbvolume, No. \vldbissue\ %
	ISSN 2150-8097. \\
	\href{https://doi.org/\vldbdoi}{doi:\vldbdoi} \\
}\addtocounter{footnote}{-1}\endgroup

\ifdefempty{\vldbavailabilityurl}{}{
	\vspace{.3cm}
	\begingroup\small\noindent\raggedright\textbf{PVLDB Artifact Availability:}\\
	The source code, data, and/or other artifacts have been made available at \url{https://github.com/rucjrliu/QSPN_code}.
	\endgroup
}

\input{secs/intro}

\input{secs/pred}

\input{secs/highlevel}
\input{secs/qspn}
\input{secs/offline}
\input{secs/online}
\input{secs/multitab}
\input{secs/expres}
\input{secs/rw}
\input{secs/conclusion}

%
%
%
\newpage
\input{reference}


\appendix

\input{secs/appendixA}

\input{secs/appendixB}
\input{secs/appendixC}

%

\end{document}

%% file: secs/commands.tex
\newcommand{\eat}[1]{}

\usepackage{latexsym}
\usepackage{amsthm}
\usepackage{xcolor}
\usepackage{colortbl}
\usepackage{epsfig}
\usepackage{xspace}
\usepackage{graphicx}
\usepackage{subfigure}
\usepackage{paralist}
\usepackage{enumerate}
\usepackage[color,matrix,arrow,all]{xy}
\usepackage{comment}
\usepackage{booktabs}
\usepackage{balance}
\usepackage{stmaryrd}
\usepackage{pifont}
\usepackage{hhline}
\usepackage{listings}
\usepackage{array}
\usepackage{float}
\usepackage[flushleft]{threeparttable}

\usepackage{mathrsfs}
\usepackage{makecell}
\usepackage{xparse}
\usepackage{wrapfig}

\usepackage{epsfig}
\usepackage{multirow}
\usepackage{url}

\usepackage{multirow}
\usepackage{natbib}
\usepackage{graphicx}

\usepackage{listings}
\usepackage{framed}
\usepackage{xcolor}
\usepackage{color}
\usepackage{geometry}
\usepackage{changepage}
\colorlet{shadecolor}{gray!20}

\definecolor{shadecolor}{RGB}{220,220,220}

\definecolor{inputcolor}{RGB}{255,139,35}
\definecolor{outputcolor}{RGB}{120,212,252}
\definecolor{embedcolor}{RGB}{254,127,156}
\definecolor{maskcolor}{RGB}{122,128,255}
\definecolor{ecolor}{RGB}{58,149,54}

\definecolor{highcolor}{RGB}{255,153,153}
\definecolor{midcolor}{RGB}{255,204,204}
\definecolor{lowcolor}{RGB}{204,229,255}

\newtheorem{example}{Example}

\usepackage{tikz}
\usetikzlibrary{shapes,snakes}
\usetikzlibrary{calc}

\usepackage[export]{adjustbox}

\definecolor{green}{RGB}{0,128,0}

\definecolor{yellow}{RGB}{255,200,18}

\newcommand{\att}[1]{\textbf{\texttt{#1}}}

\newcommand{\stab}{\vspace{1.2ex}\noindent}

\newcommand{\bi}{\begin{itemize}}
\newcommand{\ei}{\end{itemize}}

\newcommand{\be}{\begin{enumerate}}
\newcommand{\ee}{\end{enumerate}}
\newcommand{\beqn}{\begin{eqnarray*}}
\newcommand{\eeqn}{\end{eqnarray*}}

\newcommand{\stitle}[1]{\stab\noindent{\bf #1}}
\newcommand{\etitle}[1]{\vspace{1mm}\noindent{\underline{\em #1}}}
\newcommand{\ie}{{\em i.e.,}\xspace}
\newcommand{\eg}{{\em e.g.,}\xspace}

\newcommand{\term}[1]{{\tt #1}}

\newcommand{\sys}{\att{QSPN}\xspace}

\usepackage{amsmath, amsthm}
 
\newtheorem{lemma}{Lemma}[] 

\NewDocumentCommand{\nan}{ mO{} }{\textcolor{red}{\textsuperscript{\textit{Nan}}\textsf{\textbf{\small[#1]}}}}

\usepackage{algorithm}
\usepackage{algorithmicx}
\usepackage{algpseudocode}
\usepackage{amsmath} 
\floatname{algorithm}{Algorithm} 

\NewDocumentCommand{\cc}{mO{}}{\textcolor{blue}
{\textsuperscript{\textit{CC}}\textsf{\textbf{\small[#1]}}}}

\NewDocumentCommand{\fanj}{mO{}}{\textcolor{orange}
	{\textsuperscript{\textit{fanj}}\textsf{\textbf{\small[#1]}}}}

\usepackage{cases}
\usepackage{cleveref}
\usepackage{makecell}
\usepackage{bm}

\usepackage{tablefootnote}

\usepackage{arydshln}

\definecolor{darkred}{rgb}{0.75, 0.0, 0.0}
\definecolor{darkgreen}{rgb}{0.0, 0.45, 0.0}

\definecolor{mint}{rgb}{0.62, 0.89, 0.75}
\definecolor{lightrose}{rgb}{0.93, 0.26, 0.22}
\definecolor{blu}{rgb}{0.18359375, 0.4296875, 0.7265625}

\colorlet{mint5}{mint!5}
\colorlet{mint10}{mint!10}
\colorlet{mint20}{mint!20}
\colorlet{mint30}{mint!30}
\colorlet{mint40}{mint!40}
\colorlet{mint50}{mint!50}
\colorlet{mint60}{mint!60}
\colorlet{mint70}{mint!70}
\colorlet{mint80}{mint!80}
\colorlet{mint90}{mint!90}
\colorlet{mint100}{mint!100}

\colorlet{rose5}{lightrose!5}
\colorlet{rose10}{lightrose!10}
\colorlet{rose20}{lightrose!20}
\colorlet{rose21}{lightrose!21}
\colorlet{rose30}{lightrose!30}
\colorlet{rose40}{lightrose!40}
\colorlet{rose45}{lightrose!45}
\colorlet{rose50}{lightrose!50}
\colorlet{rose60}{lightrose!60}
\colorlet{rose70}{lightrose!70}
\colorlet{rose80}{lightrose!80}
\colorlet{rose90}{lightrose!90}
\colorlet{rose100}{lightrose!100}

\colorlet{green5}{green!5}
\colorlet{green10}{green!10}
\colorlet{green20}{green!20}
\colorlet{green30}{green!30}
\colorlet{green40}{green!40}
\colorlet{green50}{green!50}
\colorlet{green60}{green!60}
\colorlet{green70}{green!70}
\colorlet{green80}{green!80}
\colorlet{green90}{green!90}
\colorlet{green100}{green!100}

\colorlet{blu5}{blu!5}
\colorlet{blu10}{blu!10}
\colorlet{blu20}{blu!20}
\colorlet{blu30}{blu!30}
\colorlet{blu40}{blu!40}
\colorlet{blu50}{blu!50}
\colorlet{blu60}{blu!60}
\colorlet{blu70}{blu!70}
\colorlet{blu80}{blu!80}
\colorlet{blu90}{blu!90}
\colorlet{blu100}{blu!100}

\colorlet{orange5}{orange!5}
\colorlet{orange10}{orange!10}
\colorlet{orange20}{orange!20}
\colorlet{orange30}{orange!30}
\colorlet{orange40}{orange!40}
\colorlet{orange50}{orange!50}
\colorlet{orange60}{orange!60}
\colorlet{orange70}{orange!70}
\colorlet{orange80}{orange!80}
\colorlet{orange90}{orange!90}
\colorlet{orange100}{orange!100}

\usepackage{scalerel}
\usepackage{stackengine}
\usepackage{pgf}
\newcounter{iloop}
\newcommand\openbigstar[1][0.7]{%
  \scalerel*{%
    \stackinset{c}{-.125pt}{c}{}{\scalebox{#1}{\color{white}{$\bigstar$}}}{%
      $\bigstar$}%
  }{\bigstar}
}
\newcommand{\Stars}[1]{\ensuremath{%
\pgfmathtruncatemacro{\imax}{ifthenelse(int(#1)==#1,#1-1,#1)}%
\pgfmathsetmacro{\xrest}{0.9*(1-#1+\imax)}%
\setcounter{iloop}{0}%
\loop\stepcounter{iloop}\ifnum\value{iloop}<\the\numexpr\imax+1
\bigstar\repeat
\openbigstar[\xrest]%
\setcounter{iloop}{0}%
\loop\stepcounter{iloop}\ifnum\value{iloop}<\the\numexpr5-\imax\relax
\openbigstar[.9]\repeat}}

\newcommand*\emptycirc[1][1ex]{\tikz\draw (0,0) circle (#1);} 
\newcommand*\halfcirc[1][1ex]{%
	\begin{tikzpicture}
	\draw[fill] (0,0)-- (90:#1) arc (90:270:#1) -- cycle ;
	\draw (0,0) circle (#1);
	\end{tikzpicture}}
\newcommand*\fullcirc[1][1ex]{\tikz\fill (0,0) circle (#1);} 

%% file: secs/notations.tex
\newcommand{\Aset}{{A}\xspace}

\newcommand{\acc}{\texttt{acc}\xspace}
\newcommand{\accvec}{\att{{acc}}\xspace}
\newcommand{\ACC}{\att{{ACC}}\xspace}
\newcommand{\Card}{\texttt{Card}\xspace}
\newcommand{\Prob}{P\xspace}

\newcommand{\cardest}{CardEst\xspace}

\newcommand{\aff}{\texttt{AFF}\xspace}
\newcommand{\ipa}{\texttt{IPA}\xspace}

\newcommand{\siq}{\texttt{SIQ}\xspace}

\newcommand{\prodnode}{\texttt{Product}\xspace}
\newcommand{\qpnode}{\texttt{QProduct}\xspace}
\newcommand{\qsnode}{\texttt{QSplit}\xspace}
\newcommand{\sumnode}{\texttt{Sum}\xspace}
\newcommand{\leafnode}{\texttt{Leaf}\xspace}

\newcommand{\rdc}{\texttt{RDC}\xspace}
\newcommand{\node}{\texttt{n}\xspace}
\newcommand{\child}{\texttt{child}\xspace}
\newcommand{\simi}{\texttt{sim}\xspace}

\newcommand{\sig}{\texttt{S}\xspace}

\newcommand{\mqspn}{\att{M-QSPN}\xspace}

%% file: secs/abstract.tex
\begin{abstract}
	Cardinality estimation is a fundamental component in database systems, crucial for generating efficient execution plans. Despite advancements in learning-based cardinality estimation, existing methods may struggle to simultaneously optimize the key criteria: \emph{estimation accuracy}, \emph{inference time}, and \emph{storage overhead}, limiting their practical applicability in real-world database environments. 
	This paper introduces \sys, a \emph{unified model} that integrates both data distribution and query workload. \sys achieves high estimation accuracy by modeling data distribution using the simple yet effective Sum-Product Network (SPN) structure. To ensure low inference time and reduce storage overhead, \sys further partitions columns based on query access patterns. 
	We formalize \sys as a tree-based structure that extends SPNs by introducing two new node types: \term{QProduct} and \term{QSplit}. This paper studies the research challenges of developing efficient algorithms for the offline construction and online computation of \sys.
	%
	%
	%
	%
	We conduct extensive experiments to evaluate \sys in both single-table and multi-table cardinality estimation settings. The experimental results have demonstrated that \sys achieves superior and robust performance on the three key criteria, compared with state-of-the-art approaches.

%

\end{abstract}

%% file: secs/intro.tex
\section{Introduction}

Cardinality estimation (\cardest), which estimates the result size of an SQL query on a relational database, is a fundamental component of query optimization in database management systems (DBMSs). Traditional \cardest methods~\cite{postgres,sample1,sample2,sample3,sample4,mhist,bayes} rely on simplifying assumptions, such as column independence, often leading to substantial estimation errors. To overcome these limitations, learning-based \cardest models~\cite{mscn,lw,naru,neurocard,mspn,deepdb,flat} have emerged as state-of-the-art solutions, significantly improving accuracy by capturing complex data distributions and query patterns.

Despite the advancements in learning-based \cardest models, deploying them in real-world DBMS systems still requires balancing three key criteria~\cite{bayes-gm3,arewe-paper,volcano,propagation,flat}: \emph{estimation accuracy}, \emph{inference time}, and \emph{storage overhead}.
Data-driven approaches~\cite{naru,neurocard,deepdb,flat} leverage probabilistic models, such as Sum-Product Networks~\cite{mspn,deepdb,flat} and Deep Auto-Regressive~\cite{naru,neurocard}, to capture the joint distribution of all columns of the relational data. While these methods typically achieve high estimation accuracy, they suffer from high inference time and substantial storage overheads, particularly when handling complex data distributions.
Query-driven methods~\cite{mscn,lw}, on the other hand, train regression models that directly map SQL queries to their estimated cardinalities based on a set of training queries, bypassing the need to model data distributions. Although these methods are efficient and lightweight, they struggle with generalization, particularly when encountering queries that significantly deviate from those in the training set.
Table~\ref{tab:intro_comp} presents a comparison of existing approaches based on the three key criteria.

Given the strengths and limitations of existing methods, a promising direction is to develop a \textbf{unified model} that leverages both data and queries, aiming to achieve the three key criteria: \emph{high estimation accuracy}, \emph{low inference time}, and \emph{lightweight storage overhead}. Such an approach can overcome the weaknesses of purely data-driven or query-driven \cardest models by leveraging complementary information from both sources.

%

Although UAE~\cite{uae} also leverages both data and queries for \cardest, its primary focus is on improving estimation accuracy by training a Deep Auto-Regressive model that incorporates unsupervised losses from data and supervised losses from queries. However, UAE inherits key limitations of data-driven approaches~\cite{naru,neurocard} that rely on Deep Auto-Regressive models. Specifically, it suffers from high inference time due to the computationally expensive progressive sampling process and incurs significant storage overhead, especially for columns with large value domains. 

\input{table/intro_tabfig}

\begin{figure*}[t]
	\centering
	\hspace{-1em}
	\subfigure[Traditional SPN Model.]{
		\includegraphics[height=0.24\textwidth]{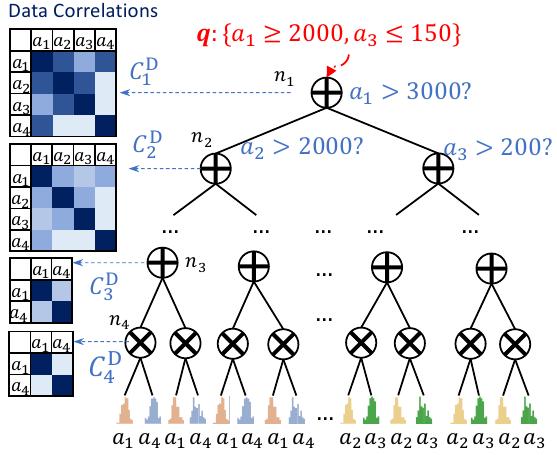}
		\label{fig:basic-idea-1}
	} 
	\subfigure[Partitioning Columns based on Query Patterns.]{
		\includegraphics[height=0.24\textwidth]{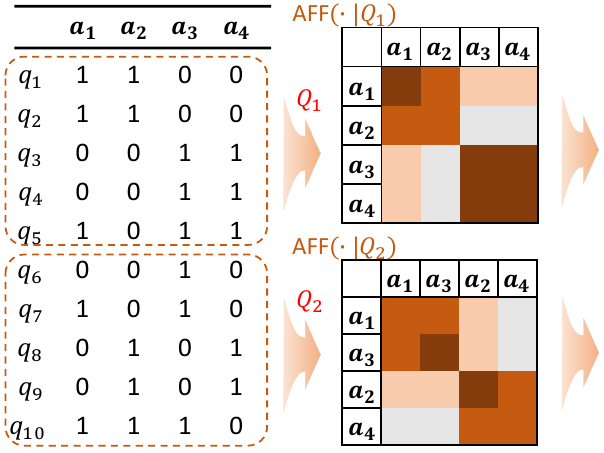}
		\label{fig:basic-idea-2}
	} 
	\subfigure[Our Proposed \sys Model]{
		\includegraphics[height=0.24\textwidth]{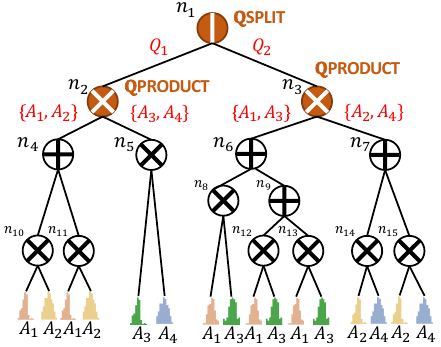}
		\label{fig:qspn-online}
	} 
	\vspace{-1em}
	\caption{High-level idea of \sys. 
	}
	\vspace{-1em}
	\label{fig:basic-idea}
\end{figure*}

\stitle{Our Proposals.} In this paper, we propose to learn from both data and queries via the Sum-Product Network (SPN) model.
As shown in Figure~\ref{fig:basic-idea-1}, traditional SPN models~\cite{deepdb,flat} recursively partition columns (\ie the \term{Product} node) and rows (\ie the \term{Sum} node) into \emph{local subsets}, making it easier to compute the joint probability distribution from these local distributions. However, they often suffer from \emph{high inference time} and \emph{large model size} when columns in a database are highly correlated. This issue arises because many intermediate nodes (\eg \term{Sum} nodes) must be introduced to ensure that the columns in partitioned subsets are treated as independent. 

To address this problem, we propose \sys that extends traditional SPNs by incorporating query workload information. The high-level idea behinds \sys stems from an observation in many real-world query workloads: queries often exhibit \textbf{specific access patterns} on the columns of relational tables, which can be effectively leveraged to enhance both the efficiency and accuracy of cardinality estimation.
%
	Take the real-world queries from the Job-Light workload~\cite{joblight} as an example which represents how users retrieve movie comments. Analyzing the query workload reveals that certain columns are frequently accessed together, while others are rarely referenced in the same queries. For instance, when retrieving movie comments by different types, production year is usually a search criteria meanwhile \ie these two columns are frequently queried together in analytical workloads, whereas company type is seldom focused together with type \ie these two columns tend to appear in a separate set of queries. Traditional SPN models overlook such query-driven correlations, leading to unnecessary model complexity and inefficiencies in inference.


By integrating query workload information, \sys can jointly partition columns based on both data correlations and query access patterns, thereby reducing model size and improving inference efficiency without sacrificing estimation accuracy, as shown in Table~\ref{tab:intro_comp}.
\begin{example}
	\label{eg:basic-idea}	
	We consider the \cardest task for an example table $T$ with highly correlated columns ${a_1, a_2, a_3, a_4}$, as illustrated in Figure~\ref{fig:basic-idea-1}. SPN partitions $T$ into different row subsets via \term{Sum} nodes (\eg node $n_1$, which partitions rows based on whether $a_1 > 3000$) to reduce column correlations within each subset. However, as depicted in the figure , when columns exhibit high correlations, the SPN requires numerous \term{Sum} nodes to break down the joint distribution into local distributions over individual columns. This leads to a substantial increase in model size. Moreover, when processing a query $q$, the inference procedure must traverse a large number of these nodes, significantly increasing inference time.
	
	As illustrated in Figure~\ref{fig:basic-idea-2}, \sys leverages the \textbf{access patterns} of the query workload $Q$, \ie how often certain columns are accessed together by the same queries. Within subset $Q_1$ of $Q$, we observe that columns $a_1$ and $a_2$ are frequently accessed together by queries $q_1$ and $q_2$, while $a_3$ and $a_4$ are jointly accessed by queries $q_3$ and $q_4$.
	Based on this pattern, we partition the columns into $\{a_1, a_2\}$ and $\{a_3, a_4\}$ for queries in $Q_1$, even if their data remain highly correlated. Similarly, for queries in subset $Q_2$, the columns can be partitioned into $\{a_1, a_3\}$ and $\{a_2, a_4\}$. These query-aware column partitions allow \sys to construct more compact SPN models, reducing model size and inference time while maintaining accuracy.
\end{example}

As shown in Figure~\ref{fig:qspn-online}, we formalize \sys as a tree-based structure that extends traditional SPNs by introducing two new node types: \term{QProduct} and \term{QSplit}.
Specifically, \term{QProduct} partitions columns based on query-specific access patterns (\eg grouping frequently accessed columns together) within a given workload, while \term{QSplit} refines the workload itself into more granular subsets to capture workload variations. Moreover, \sys retains the SPN's ability to partition columns and rows based on data correlations.
By partitioning columns based on both data correlations and query access patterns, \sys effectively reduces the number of intermediate nodes in the SPN, which improves inference efficiency, reduces storage overhead while maintaining high estimation accuracy.



\stitle{Key Challenges and Solutions.}
We study the technical challenges that naturally arise in our proposed \sys approach.

\etitle{Offline \sys Construction.}
A key challenge in offline \sys construction is integrating query workload information into the SPN framework while maintaining inference efficiency. Unlike conventional SPNs, 
\sys must consider query co-access patterns, making the partitioning problem more complex.
To address this, \sys develops efficient algorithms for two core problems, \ie \qpnode for query-aware column partitioning and \qsnode for workload partitioning.




\etitle{Online \sys Computation.} 
The online stage of \sys presents two key challenges. First, accurately computing query cardinalities while minimizing inference overhead is non-trivial, especially for queries with unseen access patterns. Second, as data distributions and query workloads evolve, \sys must maintain accuracy without requiring frequent full retraining.
To tackle these challenges, we develop an online inference algorithm, and introduce an incremental update mechanism that enables \sys to efficiently adapt to workload and data changes.

\etitle{Multi-Table \cardest with \sys.}
Extending \sys to multi-table cardinality estimation is challenging due to the complex distributions of join keys and their impact on base table query predicates. 
%
To address this, we introduce a novel approach that allows \sys to generalize to multi-table queries while maintaining high accuracy and efficient inference performance.

\stitle{Contributions.} Our contributions are summarized as follows.

\vspace{1mm} \noindent
(1) We propose \sys, a query-aware Sum-Product Network that integrates data and queries for \cardest (Section~\ref{sec:qspn}).

\vspace{0.5mm} \noindent
(2) We develop effective algorithms for offline \sys construction and online \sys computation, balancing estimation accuracy, inference time, and model size (Sections~\ref{sec:qspn-offline} and \ref{sec:qspn-online}). We extend \sys to support multi-table cardinality estimation (Section~\ref{sec:qspn-multi}). 

\vspace{0.5mm} \noindent
(3) We conduct a comprehensive experimental study on widely used \cardest benchmarks. Extensive results demonstrate that \sys achieves superior performance (Section~\ref{sec:exp}).

\vspace{-1em}

%% file: table/intro_tabfig.tex
\begin{table}[t!]
	\caption{Cardinality Estimation Methods Comparison} 
	\vspace{-2mm}
	\label{tab:intro_comp}
	\resizebox{0.98\columnwidth}{!}{
		\begin{tabular}{|c|c||c|c|c|}
			\hline
			\multirow{2}*{\textbf{Method}} & \multirow{2}*{\textbf{Category}}  & \textbf{Accurate} & \textbf{Fast} & \textbf{Lightweight} \\
			& & \textbf{Estimation} & \textbf{Inference} & \textbf{Storage} \\
			\hline
			\hline
			Postgress & Traditional  & \emptycirc & \fullcirc & \fullcirc \\
			MHist & Traditional & \emptycirc & \emptycirc & \halfcirc \\
			Sampling & Traditional & \emptycirc & \fullcirc & \emptycirc \\
			\hline
			\hline
			MSCN~\cite{mscn} & Query-Driven & \emptycirc & \fullcirc & \halfcirc \\
			LW-XGB~\cite{lw} & Query-Driven & \emptycirc & \fullcirc & \fullcirc \\
			\hline
			\hline
			Naru~\cite{naru} & Data-Driven & \fullcirc & \emptycirc & \halfcirc \\
			DeepDB~\cite{deepdb} & Data-Driven & \halfcirc & \halfcirc & \halfcirc\\
			FLAT~\cite{flat} & Data-Driven & \fullcirc & \halfcirc & \emptycirc \\
			\hline \hline
			UAE~\cite{uae} & Hybrid & \fullcirc & \emptycirc & \halfcirc \\
			\hline \hline
			\textbf{Ours} & \textbf{Hybrid} & \fullcirc & \fullcirc & \fullcirc \\
			\hline
		\end{tabular}
		\vspace{-4em}
	}
\end{table}

%% file: secs/pred.tex
\section{Preliminaries} \label{sec:preliminaries}


%

\subsection{Problem Formalization} \label{subsec:problem}



\stitle{Data.}
This paper considers a relational table $T$ with columns (or attributes) $\Aset = \{a_1, a_2, \ldots, a_{|\Aset|}\}$ and tuples $T = \{t_1, t_2, \ldots, t_{|T|}\}$, where $T$ may be either a single table or a joined table. 
Following existing work~\cite{flat}, we define the domain of each column $a_i$ as $[LB_i, UB_i]$, where $LB_i$ and $UB_i$ represent the lower and upper bounds of $a_i$, respectively.

\stitle{Queries.}
Similar to existing works~\cite{deepdb, flat, uae}, this paper focuses on queries that consist of a \emph{conjunction} of predicates, where each predicate over column $a_i$ can be represented as $L_i \leq a_i \leq U_i$, with $LB_i \leq L_i \leq U_i \leq UB_i$. Without loss of generality, the endpoints of interval $[L_i, U_i]$ can also be open, \eg $(L_i, U_i]$ or $[L_i, U_i)$, which are omitted in this paper for simplicity. In particular, a point query over $a_i$ can be represented as $L_i = U_i$. 
%
%
%
For ease of presentation, we assume each column has only one interval $[L_i, U_i]$, though this can be easily extended to the cases with multiple intervals per column.

This paper considers a \emph{query workload} as a set of queries $Q = \{q_1, q_2, \ldots, q_{|Q|} \}$, typically extracted from real query logs.
Given this workload $Q$, we introduce \emph{query-column access matrix} (or \emph{access matrix} for short) to represent the \textbf{access patterns} of queries in $Q$ on columns in $T$.
This matrix provides a structured way to capture which queries reference which columns, enabling more optimization opportunities in query-aware cardinality estimation.
\begin{definition}[Query-Column Access Matrix]
	For each query $q_i$ and each column $a_j$, we define $\acc(q_i, a_j)$ as an indicator function that specifies whether column $a_j$ is accessed by query $q_i$. Specifically, $\acc(q_i, a_j) = 1$ if $q_i$ accesses $a_j$, and $\acc(q_i, a_j) = 0$ otherwise.
	Then, we define the \emph{access matrix} for a workload $Q$ and a column set $\Aset$, denoted as $\ACC(Q, \Aset)$, as a binary matrix where each entry $\acc(q_i, A_j)$ indicates whether query $q_i$ accesses column $a_j$. 
\end{definition}
In particular, we use $\accvec(q_i)$ to represent the $i$-th row of the access matrix, corresponding to the access pattern of query $q_i$ across all columns, and $\accvec(a_j)$ to denote the $j$-th column of the access matrix, capturing how column $a_j$ is accessed by queries.

\stitle{Cardinality Estimation.}
Given a new query $q$, we define $\Card(q, T)$ as the \emph{cardinality} of $q$ over table $T$, \ie the number of tuples in $T$ that satisfy the query conditions. 
The goal of \emph{cardinality estimation} is to compute an estimate $\widehat{\Card}(q, T)$ that approximates $\Card(q, T)$ efficiently and accurately without executing $q$ on $T$.
Specifically, we study the following problems: (1) \emph{Offline \cardest Training}, which trains a \cardest model by leveraging both data $T$ and workload $Q$, capturing the underlying data distribution and query access patterns, and (2) \emph{Online \cardest Inference}: which uses the trained \cardest model to estimate the cardinality $\widehat{\Card}(q, T)$ for a given query $q$. 

\subsection{SUM-Product Network (SPN) Model} \label{subsec:spn}

Sum-Product Network (SPN)~\cite{spn} is a data-driven model with extensions such as DeepDB~\cite{deepdb} and FLAT~\cite{flat}. SPN-based approaches address the \cardest problem by modeling the \emph{joint probability distribution} $\Prob_{T}(\Aset)$, where each attribute $a_i \in \Aset$ is treated as a random variable.
Given a query $q$ with selection predicates ${A_i \in [L_i, U_i]}_{i=1}^{m}$, the estimated cardinality is computed as
$
	\widehat{\Card}(q, T) =|T| \cdot \sum_{v_1 \in [L_1, U_1]} \ldots \sum_{v_m \in [L_m, U_m]} {\Prob_{T}(v_1, \ldots, v_m)},
$
where the summation iterates over all possible values within the query's range constraints. 
%
SPN approximates $\Prob_{T}(\Aset)$ by \emph{decomposing} the joint probability distribution into multiple \emph{local probability distributions}. This decomposition is realized through a hierarchical, tree-based structure, where each node represents a \emph{local} joint probability distribution $\Prob_{T^\prime}(\Aset^\prime)$.


The key idea of SPN focuses on introducing intermediate nodes, which fall into one of the following two categories.
(1) A \term{Sum} node partitions its tuple set $T^\prime$ into a collection of disjoint subsets $T^\prime = \bigcup_{i}{T_{i}^\prime}$. Each subset $T_{i}^\prime$ corresponds to a child node with a probability distribution $\Prob_{T_{i}^\prime}(\Aset^\prime)$. The overall distribution at the \term{Sum} node is then computed as a weighted sum of its children's distributions
	$
		\Prob_{T^\prime}(\Aset^\prime) = \sum_{i} w_i \cdot \Prob_{T_{i}^\prime}(\Aset^\prime),
	$
	where the weight $w_i$ is determined by the proportion of tuples in each subset, given by $w_i = |T_{i}^\prime| / |T^\prime|$.
(2) A \term{Product} node partitions its attribute set $\Aset^\prime$ into disjoint subsets $\Aset^\prime = \bigcup_{j}{\Aset_{j}^\prime}$. Each subset $\Aset_{j}^\prime$ corresponds to a child node that models the probability distribution $\Prob_{T^\prime}(\Aset_{j}^\prime)$. By assuming independence among these subsets, the overall distribution at the \term{Product} node is computed as
	$
		\Prob_{T^\prime}(\Aset^\prime) = \prod_{j}{\Prob_{T^\prime}(\Aset_{j}^\prime)}
	$
	This decomposition allows SPNs to efficiently approximate complex joint distributions by leveraging independence between attributes in a particular data subset.
Given a query $q$ for cardinality estimation, SPN estimates the cardinality $\widehat{\Card}(q, T)$ in a bottom-up manner.


\vspace{-1em}

%% file: secs/highlevel.tex
\section{An Overview of \sys} \label{sec:qspn}

%% file: secs/qspn.tex
We propose \sys that extends traditional SPNs by incorporating query workload information to partition columns based on their \emph{access patterns}. We formalize \sys as a tree-based structure that extends traditional SPNs by introducing two new node types, \qpnode and \qsnode, as shown in Figure~\ref{fig:qspn-online}.
%
%
%
Formally, each node $\node$ in \sys is represented as a $4$-tuple $(\Aset_{\node}, T_{\node}, Q_{\node}, O_{\node})$, where $\Aset_{\node}$ denotes the column set, $T_{\node}$ the corresponding table, and $Q_{\node}$ the associated query workload. Each node captures the joint probability distribution $P_{T_{\node}}(\Aset_{\node})$ conditioned on the queries in $Q_{\node}$. Moreover, the node type $O_{\node}$ represents how the joint probability of node $\node$ is estimated from its child nodes, which is described as follows. 


\stitle{\qpnode.}
A \qpnode node, such as $\node_2$ in Figure~\ref{fig:qspn-online}, partitions its column set $\Aset_{\node}$ into a set of disjoint subsets $\mathcal{A} = \{\Aset_1, \Aset_2, \ldots, \Aset_m\}$, ensuring that columns in different subsets are infrequently co-accessed by queries in $Q_{\node}$.
For each subset $\Aset_i$, a corresponding child node is created as $\node.\child_i = (\Aset_i, T_{\node}[\Aset_i], Q_{\node}[\Aset_i], O_{i})$. The joint probability distribution $\Prob_{T_{\node}}(\Aset_{\node})$ at node $\node$ can then be computed as
$
	\Prob_{T_{\node}}(\Aset_{\node}) = \prod_{i=1}^{m}{\Prob_{T_{\node}}(\Aset_{i})}.
$

\stitle{\qsnode.}
A \qsnode, such as $\node_1$ in Figure~\ref{fig:qspn-online}, splits its query workload $Q_{\node}$ into a set of disjoint query subsets $\mathcal{Q} = \{Q_1, Q_2, \ldots, Q_m\}$, ensuring that queries within the same subset share similar access patterns, while queries across different subsets exhibit distinct access behaviors.
For each query subset $Q_i$, a corresponding child node is created as $\node.\child_i = (\Aset_{\node}, T_{\node}, Q_{i}, O_{i})$.
Note that a \qsnode does not directly compute a joint probability distribution $P_{T_{\node}}(\Aset_{\node})$ but instead functions as a \emph{query router}. Specifically, when estimating the cardinality of a query $q$, the \qsnode identifies the query subset $Q_{k^{*}}$ from $\mathcal{Q}$ that shares the most \emph{similar access patterns} with $q$ and routes $q$ to the corresponding child node $\node.\child_{k^{*}}$ for cardinality estimation.
We will discuss the method for determining whether a query subset exhibits the most \emph{similar access patterns} with $q$ later.

\stitle{\prodnode.} 
A \prodnode node, such as $\node_5$ in Figure~\ref{fig:qspn-online}, partitions its column set $\Aset_{\node}$ into a set of disjoint subsets $\mathcal{A} = \{\Aset_1, \Aset_2, \ldots, \Aset_m\}$, ensuring statistical independence between columns in different subsets. 
For each subset $\Aset_i$, a corresponding child node is created as $\node.\child_i = (\Aset_i, T_{\node}, Q_{\node}, O_{i})$. The joint distribution $\Prob_{T_{\node}}(\Aset_{\node})$ is then computed using the equation in Section~\ref{subsec:spn}.

\stitle{\sumnode.}
A \sumnode node, such as $\node_4$ in Figure~\ref{fig:qspn-online}, partitions the table $T_{\node}$ of node $\node$ into disjoint subsets $\mathcal{T} = \{T_1, T_2, \ldots, T_{m}\}$. For each subset $T_i$, a corresponding child node is created as $\node.\child_i = (\Aset_{\node}, T_{i}, Q_{\node}, O_{i})$. The joint distribution $\Prob_{T_{\node}}(\Aset_{\node})$ is then computed as a weighted sum of the distributions of its child nodes, as defined in Section~\ref{subsec:spn}.

\stitle{\leafnode.}
A \leafnode node, such as any leaf in the tree shown in Figure~\ref{fig:qspn-online}, represents the 1-dimensional probability distribution $\Prob_{T_{\node}}(\Aset_{\node})$. Specifically, we use a histogram-based mechanism to capture this probability distribution. The construction cost of the histogram is $O(|T_{\node}|)$, and the query cost is approximately $O(1)$.

\vspace{1mm}

To support the above \sys structure, in this paper, we introduce a framework that consists of both offline and online stages. 

\stitle{Offline \sys Construction.}
In the offline stage, \sys learns its structure from both data $T$ and workload $Q$, capturing the underlying data distribution and query access patterns. Unlike conventional SPNs, \sys introduces a new challenge of incorporating query co-access patterns into column partitioning. 
We propose efficient algorithms \qpnode and \qsnode, as presented in Section~\ref{sec:qspn-offline}.

\stitle{Online \sys Computation.}
In the online stage, we utilize \sys for cardinality estimation. Specifically, for queries with access patterns that differ from those in the training workload, we develop an online inference algorithm, ensuring both accuracy and efficiency. 
%
%
Second, we introduce an incremental update mechanism that selectively identifies and updates only the affected parts of the model.
%
%
Details of online \sys inference are provided in Section~\ref{sec:qspn-online}.

\stitle{Multi-Table Cardinality Estimation.}
This paper also explores extending \sys to support multi-table \cardest. The key difficulty lies in accurately modeling join key distributions while effectively handling multi-table query predicates. 
Traditional methods rely on heuristic bucket-based approaches, which suffer from poor accuracy. Learning-based \cardest models~\cite{flat,neurocard}, on the other hand, train on a materialized outer-join table, incurs excessive time and storage costs.
To tackle this, we introduce an effective approach, which is described in Section~\ref{sec:qspn-multi}.

%% file: secs/offline.tex
\section{Offline QSPN Construction}\label{sec:qspn-offline}

Given a relational table $T$ with column set $\Aset$ and a query workload $Q$, \sys construction generates a \sys tree that models the joint probability distribution $\Prob_{T}(\Aset)$ conditioned on $Q$. To achieve this, the construction process recursively decomposes the joint probability distribution into local probability distributions in a top-down manner.
%
Specifically, during the construction of each node, \sys attempts different node types in the following order: \leafnode, \prodnode, \qpnode, \qsnode, and \sumnode. 

%


\stitle{Construction of \leafnode Nodes.}
During the recursive process, if $\Aset$ contains only a single column, this indicates that the joint probability distribution has been fully decomposed into a local distribution over the specific column. 
In this case, the construction process creates a \leafnode to model the 1-dimensional probability distribution $P_{T_n}(\Aset)$ using a histogram. This choice is motivated by the histogram's accuracy, efficiency, and lightweight nature, making it well-suited for modeling such distributions.

\stitle{Construction of \prodnode Nodes.} A \prodnode node is constructed when the column set $\Aset$ of a node exhibits statistical dependencies suitable for partitioning. Following prior works~\cite{spn,deepdb,flat}, we use the \emph{Randomized Dependence Coefficient} (RDC) to measure statistical dependencies between columns. Details on RDC can be found in the original paper~\cite{DBLP:conf/nips/Lopez-PazHS13}. Then, we employ a partition-based method. Specifically, using these RDC values, we construct a graph where vertices represent columns and edges represent dependencies weighted by the RDC values. Then, we remove the edges with RDC values below a threshold, and divide the graph into connected components, each representing an independent subset of columns.

\stitle{Construction of \sumnode Nodes.}
During the recursive process, if all other types of nodes fail to meet the decomposition criteria, node $n$ defaults to a \sumnode, which splits the data $T$ into subsets ${T_{i}}$. To achieve this, following prior works~\cite{deepdb,flat}, we use the K-Means clustering algorithm, as it partitions the data into clusters, which helps to reduce data correlation within each subset.


%

\subsection{Construction of \qpnode}
\label{subsec:construct-qproduct}

\qpnode partitions columns according to their query access patterns, grouping frequently co-accessed columns together while separating those that are rarely co-accessed into distinct subsets.

\stitle{Formalization of \qpnode.}
We first formally define \emph{access affinity} using the query-column access matrix as follows.
\begin{definition}[Access Affinity]
	The \emph{Access Affinity} between columns $a_i$ and $a_j$ with respect to query workload $Q$, denoted by $\aff(a_i, a_j | Q)$, is defined as how frequently both columns are referenced together by queries in $Q$, \ie
	\begin{equation}
		\aff(a_i, a_j | Q) = \accvec(a_i)\cdot\accvec(a_j).
	\end{equation}
\end{definition}

Using access affinity, we formally define the \qpnode operation.
\begin{definition}[\qpnode]
	Given a query workload $Q$ and a column set $\Aset = \{a_1, a_2, \ldots, a_{|\Aset|}\}$, \qpnode partitions $\Aset$ into a set of disjoint column subsets $\mathcal{A} =\{A_1, A_2, \ldots, A_m\}$, minimizing the \emph{inter-partition affinity} (\emph{IPA}) of partitioning $\ipa(\mathcal{A}|Q)$, where
	\begin{equation} \label{eq:ipa}
		\ipa(\mathcal{A} | Q) = \sum_{1 \leq k < l \leq m}{\sum_{a_i \in A_k} \sum_{a_j \in A_l} {\aff(a_i, a_j | Q)}}.
	\end{equation}
\end{definition}

For example, consider the workload $Q_1$ in Figure~\ref{fig:basic-idea-2}. The inter-partition affinity for the column partitioning $\{\{a_1, a_2\}, \{a_3, a_4\}\}$ is $0$, whereas for the partitioning $\{\{a_1, a_3\}, \{a_2, a_4\}\}$ it is $4$. Based on these results, \qpnode selects the partition $\{\{a_1, a_2\}, \{a_3, a_4\}\}$ for workload $Q_1$ to minimize inter-partition affinity.


\stitle{Algorithm Design for \qpnode.}
We first analyze the complexity of the \qpnode construction problem, as shown below.
\begin{lemma}
	The problem of \qpnode construction is equivalent to the minimum $k$-cut problem. 
\end{lemma}

We omit the proof due to the space constraint.
The minimum $k$-cut problem, even for fixed $k$, is computationally expensive to solve. We abstract each column to a vertice. For example, the minimum $2$-cut (\ie the classic min-cut problem) can be solved in $O(|A|^3)$ time using algorithms such as Stoer-Wagner~\cite{2cut}. For $k = 3$, the time complexity is $O(|A|^3\tau(|A|))$~\cite{3cut}, where $\tau(|A|)$ represents the cost of computing the objective function, which is $O(|A|^2)$ in our \qpnode construction problem, yielding an overall complexity of $O(|A|^5)$. Similarly, a recent algorithm for $k = 4$ achieves a complexity of $O(|A|^6\tau(|A|))$ (or $O(|A|^8)$). While the problem is polynomial-time solvable for fixed $k \ge 5$, the complexity increases dramatically (\eg $O(|A|^{16})$ for a minimum 5-cut algorithm as suggested by~\cite{4cut}), rendering such algorithms impractical for real-world use.

Given the computational expense of partitioning the column set to minimize IPA, we design an algorithm \att{PartitionByAFF} that achieves effective and efficient ($O(|A|^2)$) results. The algorithm first constructs a graph $G = (A, E)$, where vertices $A$ represent columns, and an edge $e_{ij} \in E$ connects columns $a_i$ and $a_j$ if their affinity $\aff(a_i, a_j | Q)$ exceeds a threshold $\tau$. Next, the algorithm identifies the connected components of $G$, with the vertices in each connected component $G_i \subseteq G$ forming a column partition $A_i$.

\subsection{Construction of \qsnode} \label{subsec:construct-qsplit}

When considering the entire workload $Q$, \qpnode may struggle to derive a meaningful column partitioning with a sufficiently low IPA score due to the presence of queries exhibiting diverse access patterns. The following example illustrates this challenge.
\begin{example}
	Given the workload $Q=\{q_1, q_2, \ldots, q_{10}\}$, as shown in Figure~\ref{fig:basic-idea-2}, an optimal column partitioning for \qpnode with $k=2$ results in $\{\{a_1, a_2, a_3\},\{a_4\}\}$, yielding a minimum IPA score of 6, which remains relatively high. The primary reason for this is that different subsets of queries within $Q$ exhibit different access patterns, leading to conflicting preferences for column partitioning. 
	Specifically, queries in $Q_1 \subset Q$ favor the column partitioning $\{\{a_1, a_2\}, \{a_3, a_4\}\}$, while those in $Q_2 \subset Q$ prefer $\{\{a_1, a_3\}, \{a_2, a_4\}\}$. These conflicting preferences arise due to the distinct access patterns exhibited by queries in different subsets. 
	%
\end{example}
%
%
\stitle{Formalization of \qsnode.}
To address the above challenge, we introduce the \qsnode operation, which partitions the query workload into $n$ subsets, \ie $\mathcal{Q} = \{Q_1, \ldots, Q_n\}$, ensuring that each subset exhibits more consistent access patterns, thus enabling \qpnode to derive more meaningful column partitions. 
%
\begin{definition}[\qsnode]
	Given a query workload $Q$ and a column set $\Aset = \{a_1, a_2, \ldots, a_{|\Aset|}\}$, \qsnode aims to partition $Q$ into a set of disjoint query  subsets $\mathcal{Q} =\{Q_1, Q_2, \ldots, Q_n\}$, minimizing the objective
$
	\sum_{k=1}^{n} {\ipa(\mathcal{A}^{*}_{k} | Q_k)},
$
where $\mathcal{A}^{*}_{k}$ is the \emph{optimal} column partition given query subset $Q_k$.
\end{definition}

For example, by splitting the workload $Q$ in Figure~\ref{fig:basic-idea-2} into two subsets ($n=2$), an effective partitioning results in $Q_1=\{q_1, \ldots, q_5\}$ and $Q_2=\{q_6, \ldots, q_{10}\}$. Given these subsets, the optimal \qpnode column partitioning for $Q_1$ is $\mathcal{A}^{*}_{1} = \{\{a_1, a_2\}, \{a_3, a_4\}\}$ with $\ipa(\mathcal{A}^{*}_{1} | Q_1) = 2$. Similarly, for $Q_2$, the optimal partitioning is $\mathcal{A}^{*}_{2} = \{\{a_1, a_3\}, \{a_2, a_4\}\}$ with $\ipa(\mathcal{A}^{*}_{2} | Q_2) = 2$. 

\stitle{Algorithm Design for \qsnode.}
The \qsnode construction problem is highly challenging because it requires simultaneously minimizing $\ipa(\mathcal{A}^{*}_{k} | Q_k) $ for each partitioned subset $Q_k$. 
Thus, instead of directly optimizing $\ipa(\mathcal{A}^{*}_{k} | Q_k)$ for each $Q_k$, we focus on minimizing its upper bound, denoted as $\overline{\ipa}(\mathcal{A}_{k} | Q_k)$, which allows us to design a more tractable algorithm while still ensuring effective workload partitioning.
Formally, the objective is to partition $Q$ into $\mathcal{Q} = \{Q_1, Q_2, \ldots, Q_n\}$ while minimizing the upper bound of inter-partition affinity, \ie $\sum_{k=1}^{n} {\overline{\ipa}(\mathcal{A}_{k} | Q_k)}$.

Next, we first design a mechanism to construct the upper bound $\overline{\ipa}(\mathcal{A}_{k} | Q_k)$ and prove that optimizing this upper bound is NP-hard. Finally, we propose an efficient heuristic algorithm that achieves effective partitioning results in practice.

\begin{figure}[t]
	\centering
	\subfigure[Upper bound design.]{
		\includegraphics[width=0.47\columnwidth]{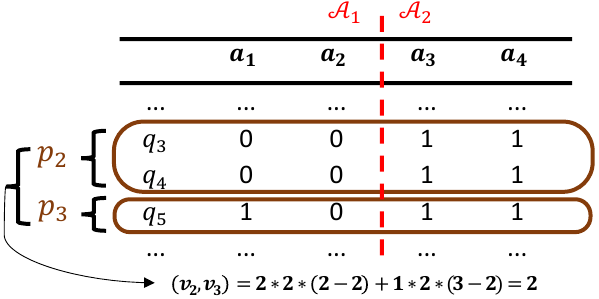}
		\label{fig:SingleTab_QS_a}
	} 
	\subfigure[Our greedy algorithm.]{
		\vspace{-2em} 
		\includegraphics[width=0.46\columnwidth]{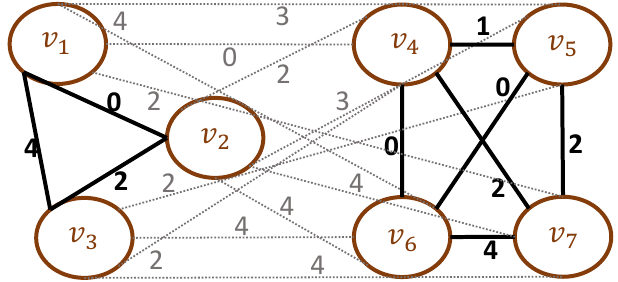}
		\label{fig:SingleTab_QS_b}
	} 
	\vspace{-1em}
	\caption{An example of \qsnode construction.}
	\vspace{-1em}
	\label{fig:SingleTab_QS}
\end{figure}

\etitle{Upper-bound design.}
Given a specific query set $Q$, we construct an upper bound $\overline{\ipa}(\mathcal{A} | Q)$ for the optimal column partitioning result $\ipa(\mathcal{A}^{*} | Q)$ by analyzing the access patterns of queries in $Q$.
Since different queries (\eg $q_1$ and $q_2$ in Figure~\ref{fig:basic-idea-2}) may share the same access pattern (e.g., $(1,1,0,0)$), we introduce $p_i$ to denote the $i$-th distinct access pattern in $Q$ and let $n_i$ represent the number of queries in $Q$ that follow pattern $p_l$. 
For instance, in Figure~\ref{fig:basic-idea-2}, the first access pattern in $Q$ is $p_1 = (1,1,0,0)$, which corresponds to two queries, \ie $n_1 = 2$. Based on these notations, we construct the following upper bound for $\ipa(\mathcal{A}^{*} | Q)$.
\begin{definition}[Upper Bound] \label{def:upper-bound}
	Given a query set $Q$ consisting of $m$ distinct query patterns $\{p_1, p_2, \dots, p_m\}$, let $\| p_i \|$ denote the $L_2 $ norm of pattern $p_i$, $n_i$) represent the number of queries in $Q$ corresponding to pattern $p_i$, and $z_{ij}$ denote the dot product of two patterns $p_i$ and $p_j$. We construct an upper bound for $\ipa(\mathcal{A}^{*} | Q)$ as
	\begin{equation} \nonumber
		\overline{\ipa}(\mathcal{A} | Q) = \sum_{i < j} \left(
		n_i \cdot z_{ij} \cdot  (\| p_i \| - z_{ij}) + n_j \cdot z_{ij} \cdot  (\| p_j \| - z_{ij})
		\right).
	\end{equation}
\end{definition}
%
Consider a query set $Q$ with two query patterns, $p_2$ and $p_3$, as illustrated in Figure~\ref{fig:SingleTab_QS_a}. To compute the upper bound $\overline{\ipa}(\mathcal{A} | Q)$, we first calculate the dot product $z_{23} = p_2 \cdot p_3 = 2$, as well as the norms $\|p_2\| = 2$ and $\|p_3\| = 3$. Given that $n_2 = 2$ and $n_3 = 1$, the contribution of patterns \( p_2 \) and \( p_3 \) to the overall upper bound is computed as:
$
n_2 \cdot z_{23} \cdot (\| p_2 \| - z_{23}) + n_3 \cdot z_{23} \cdot (\| p_3 \| - z_{23}) = 2.
$
As shown in Figure~\ref{fig:SingleTab_QS_a}, the upper bound corresponds to a specific column partitioning strategy, \ie grouping columns that are co-accessed by multiple query patterns (\eg $A_3$ and $A_2$) while placing the remaining columns (\eg $A_1$ and $A_4$) in a separate group.
%
\begin{lemma}
	For any given query set \( Q \), the upper bound \( \overline{\ipa}(\mathcal{A} | Q) \) provides an upper estimate of the optimal result \( \ipa(\mathcal{A}^{*} | Q) \).
\end{lemma}

Due to the space limit, we omit the proof in this paper.

\etitle{Hardness of upper-bound optimization.}
Next, we show that even optimizing the upper bound \(\overline{\ipa}(\mathcal{A} | Q)\) is theoretically intractable.
\begin{lemma}
The problem of partitioning \( Q \) into \( \mathcal{Q} = \{Q_1, Q_2, \ldots, Q_n\} \) to minimize the upper bound of inter-partition affinity, \ie, \( \sum_{k=1}^{n} {\overline{\ipa}(\mathcal{A}_{k} | Q_k)} \), is NP-hard.
\end{lemma}

We prove this lemma via a reduction from the Max-Cut problem, which is NP-hard. Due to space constraints, we omit the proof.

%


\etitle{A greedy algorithm for \qsnode.}
Given the time complexity of solving our problem, we design a practical and efficient heuristic algorithm.
The algorithm first constructs a graph $G = (V, E)$, where each vertex $v_i \in V$ represents a query pattern $p_i$, and an edge $e_{ij} \in E$ connects two patterns $p_i$ and $p_j$ with a weight of $n_i \cdot z_{ij} \cdot (\| p_i \| - z_{ij}) + n_j \cdot z_{ij} \cdot (\| p_j \| - z_{ij})$.
Next, based on the constructed graph $G$, the algorithm aims to partition it into $n$ sub-graphs, denoted as $\{G_1, G_2, \ldots, G_n\}$, such that the total weight of the edges \emph{across} the sub-graphs is maximized. To achieve this, it first sorts all vertices in $G$ based on their weighted degree in descending order and initializes $n$ empty sub-graphs.
Then, the algorithm iteratively processes each vertex and assigns it to an \emph{appropriate} sub-graph. Specifically, in the $i$-th iteration, the algorithm considers vertex $v_i$ and evaluates its potential assignment to each sub-graph $G_j$ by computing the total edge weight of $G_j^\prime$ after incorporating $v_i$. The vertex $v_i$ is then assigned to the sub-graph that results in the minimum weight summation.
Figure~\ref{fig:SingleTab_QS_b} illustrates an example of our greedy algorithm partitioning the query set $Q$ from Figure~\ref{fig:basic-idea-2} into two subsets (\ie $m=2$).

%% file: secs/online.tex
\section{Online QSPN Computation} \label{sec:qspn-online}


\begin{figure}
	\centering
	\includegraphics[width=0.7\columnwidth]{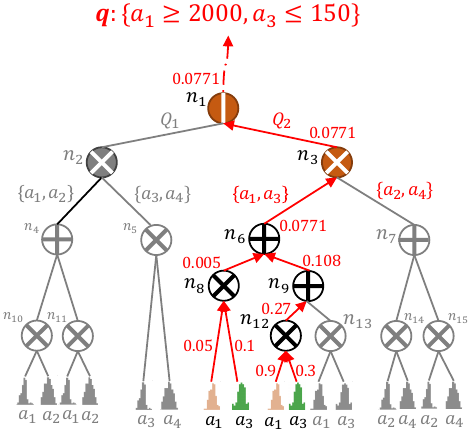}
	\vspace{-1em}
	\caption{Illustration of \cardest Inference with \sys.}
	\vspace{-2em}
	\label{fig:qspn-online-new}
\end{figure}

\subsection{\cardest Inference with \sys} \label{subsec:qspn-online}
When a new query $q$ arrives, the Online \cardest Inference process using \sys operates recursively by traversing the \sys tree from the root to the leaf nodes, as shown in Figure~\ref{fig:qspn-online-new}. Specifically, during traversal, if a \qsnode is visited, the process routes query $q$ to the child node corresponding to the most relevant query subset. 
%
On the other hand, if a \qpnode, \prodnode, or \sumnode is visited, the process computes the joint probability accordingly as follows. 

(1) For a \leafnode node $\node$, the algorithm  computes the probability	$\node.P_T(q)$ based on the histogram at leaf $\node$.

(2) For a \qpnode or \prodnode node, $\node$, the algorithm recursively invokes the \cardest inference process for each child of $\node$ where $q.A \cap \node.\texttt{child}_i.A \neq \emptyset$. It then multiplies the estimated probabilities of all the relevant child nodes to produce the estimation result. For example, at \qpnode node $\node_3$, the column set is partitioned into ${a_1, a_3}$ and ${a_2, a_4}$. Since $q$ involves only ${a_1, a_3}$, the estimation continues with node $\node_6$, while node $\node_7$ is pruned. 

(3) For a \sumnode node, $\node$, the algorithm computes the weighted sum of the estimated probabilities of its child nodes. 

Next, we explore two key implementation details of the \cardest Inference process: (1) query routing in \qsnode nodes, and (2) sub-tree pruning in \sumnode and \prodnode nodes.

\stitle{Query Routing in \qsnode Nodes.}
The objective of query routing in a \qsnode node (say $\node_1$ in Figure~\ref{fig:qspn-online-new}) is to measure the degree to which a query set $Q$ shares \emph{similar access patterns} with a given query $q$, which guides the \qsnode node in routing $ q $ to the appropriate child node (\eg node $\node_3$). To this end, we formally introduce a matching score $\sig(Q,q)$ between a query set $Q$ and a query $q$:
\begin{equation}
	\sig(Q,q) = \frac{1}{|Q|} \sum_{(a_i, a_j), i<j} \aff(a_i, a_j | Q) \cdot \aff(a_i, a_j | \{q\})
\end{equation}

The intuition behind the matching score $ \sig(Q, q) $ is to measure how closely the access patterns of queries in $ Q $ align with the access pattern of $ q $. For example, consider the query $ q $ shown in Figure~\ref{fig:qspn-online-new}. The access pattern for $ q $ is represented as: 
$$\aff(\cdot, \cdot | \{q\}) = [[1,0,1,0],[0,0,0,0],[1,0,1,0],[0,0,0,0]].$$

Considering the workload partitioning in Figure~\ref{fig:basic-idea-2}, we have $\aff(\cdot, \cdot | Q_1) = [[3,2,1,1],[2,2,0,0],[1,0,3,3],[1,0,3,3]]$ and $\aff(\cdot, \cdot | Q_2) = [[2,1,2,0],[1,3,1,2],[2,1,3,0],[0,2,0,2]]$ for $Q_1$ and $Q_2$, respectively.
Based on these, we compute $ \sig(Q_1, q) = \frac{1}{5} $ and $ \sig(Q_2, q) = \frac{2}{5} $. Therefore, when the node $ \node_1 $ is visited, the algorithm routes query $ q $ to its child node $ \node_3 $, as $ Q_2 $, corresponding to $ \node_3 $, shares more similar access patterns with $ q $ than $ Q_1 $. 

\stitle{Pruning Rules in \sumnode and \prodnode Nodes.}
We employ pruning rules that leverage query $q$ and pre-computed metadata to exclude irrelevant child nodes of a given node $\node$.
%
(1) For \prodnode and \qpnode nodes, let $\Aset_{q}$ denote the set of columns constrained by the query $q$. For a child node $\node.\child_{k}$ of $\node$, corresponding to the column subset $\Aset_{k}$, pruning occurs if $\Aset_{k} \cap \Aset_{q} = \emptyset$. (2) For \sumnode nodes, it decides which child nodes contributes to $n.P_T(q)$ so that participate the computation \ie set of visited child nods $\mathcal{C}$. Consider a child node $\node.\child_{k}$ of $\node$, corresponding to the table subset $T_{k}$. Before query processing, we pre-compute and store the range of values for each column $A_i$ in $T_{k}$. During cardinality estimation, if the value ranges of $T_{k}$ for any column do not overlap with the constraints specified by query $q$, the child node $\node.\child_{k}$ can be safely pruned, as it does not contribute to the result.


\subsection{\sys Model Update} \label{subsec:qspn-update}

Data updates ($ \Delta T$, which include new tuples) and query workload shifts ($ \Delta Q $, which include new queries) impact the accuracy and inference efficiency of the original \sys model.
The high-level idea of the update method is to traverse \sys in a top-down manner. Each time a node $ \node $ is visited during the traversal, two steps are performed to update the subtree rooted at $ \node $. First, the method examines whether $ \node $, originally constructed using $ \node.T $ and $ \node.Q $, still fits $ \node.T \cup \Delta T $ or $ \node.Q \cup \Delta Q $. Second, if $ \node $ no longer fits, the subtree rooted at $ \node $ is reconstructed by calling the \sys construction method (see Section~\ref{sec:qspn-offline}), which generates a new subtree rooted at $ \node^\prime$. Otherwise, each child node $ \node.child_i $ is recursively updated. Note that the check and reconstruction steps are unnecessary if $ \node $ is a \leafnode, as histograms can be incrementally updated in $ O(|\Delta T|) $.


The key challenge in the above update method is efficiently examining {whether a node \( \node \) still fits \( \node.T \cup \Delta T \) or \( \node.Q \cup \Delta Q \)}, as the corresponding data table \( \node.T \) and query workload \( \node.Q \) are not materialized at node \( \node \). To address this challenge, we maintain lightweight data structures in different types of nodes and design a mechanism to check whether node \( \node \) is still up-to-date with respect to the data and query workload, as described below.


(1) If \( \node \) is a \prodnode node, we examine whether the column partitioning still holds, \ie whether columns in different partitions remain independent with respect to the updated data table \( T \cup \Delta T \). To do this, we first compute \( \rdc(a_i, a_j | \Delta T) \), where \( a_i, a_j \in \node.A \), and then check whether there exist \( a_i, a_j \) from different child nodes, say $a_i$ from \( \node.child_k \) and $a_j$ from \( \node.child_l \) such that
$
\frac{|\node.T|}{|\node.T| + |\Delta T|} \rdc(a_i, a_j | \node.T) + \frac{|\Delta T|}{|\node.T| + |\Delta T|} \rdc(a_i, a_j | \Delta T) 
$
is larger than a pre-defined threshold. 
If any such pair \( a_i, a_j \) is found, 
the subtree rooted at \( \node \) needs to be reconstructed for more accurate \cardest. 
If the RDC between columns within a child node becomes less significant due to \( \Delta T \), we may reconstruct the subtree rooted at the child node to further partition the now-independent columns. 

(2) If \( \node \) is a \qpnode node, the update examination strategy is similar to that of the \prodnode case, except that we consider the access affinity \( \aff(a_i, a_j | \node.Q) \) for any column pair \( (a_i, a_j) \), instead of the correlation \( \rdc(a_i, a_j | \node.T) \).

(3) If \( \node \) is a \qsnode, we examine whether the query routing strategy still holds for the updated workload \( Q \cup \Delta Q \). To do this, for each workload partition \( Q_k \) corresponding to the child node \( \node.child_k \), we maintain the average of the matching scores of the queries in \( Q \) routed to \( Q_k \), i.e., \( {\sum_{q \in Q_k} \sig(Q_k, q)}/{|Q_k|} \). Then, for each query \( q' \) in the updated workload \( \Delta Q \), we assign \( q' \) to the child node with the maximum matching score (see Section~\ref{subsec:qspn-online}) and update the average matching score accordingly. If the average matching score of any workload partition becomes less significant, \eg less than a predefined threshold, we reconstruct the subtree rooted at the \qsnode node \( \node \), as the workload partition no longer reflects the access patterns of \( Q \cup \Delta Q \).

(4) If \( \node \) is a \sumnode node, we maintain the centroid for each tuple subset \( T_i \) from \( \mathcal{T} = \{T_1, T_2, \ldots, T_m\} \) and the average distance between each tuple and the centroid of its assigned subset. Then, for each new tuple \( t \) in \( \Delta T \), we assign \( t \) to the tuple subset with the minimum distance to the centroid of that subset and update the average distance accordingly. If the average distance becomes significant, \eg exceeding a predefined threshold, we update the subtree rooted at \( \node \), as the \sumnode may no longer hold for \( \node.T \cup \Delta T \).

In this way, the \sys update method minimizes unnecessary costs associated with model updates while ensuring accuracy in response to data updates and query workload shifts.

%% file: secs/multitab.tex
\section{Multi-Table \cardest with QSPN} \label{sec:qspn-multi}


\begin{figure}[t]
	\centering
	\subfigure[Multi-Table \cardest using Join-Key Binning]{
		\includegraphics[width=0.95\columnwidth]{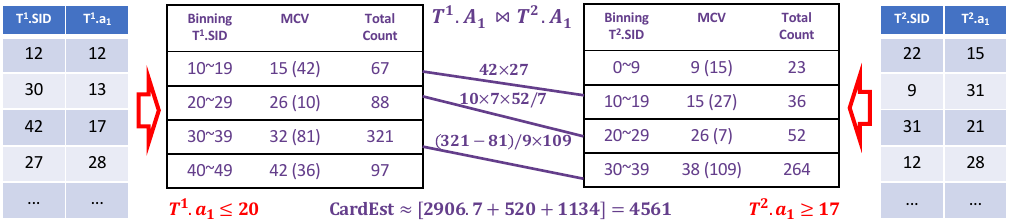}
		\label{fig:MultiTab_b}
	} 
	\subfigure[Binning Join-Key by \sys Model on a Single Table.]{
	\vspace{-2em} 
	\includegraphics[width=0.95\columnwidth]{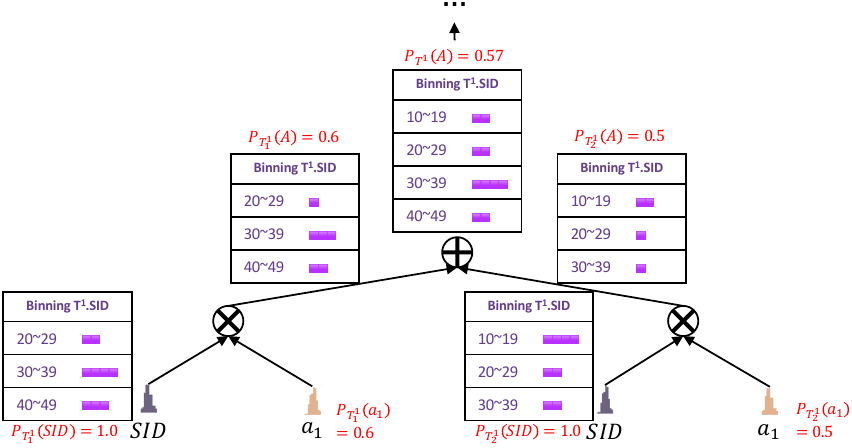}
	\label{fig:MultiTab_a}
} 
	\vspace{-1em}
	\caption{
		Illustration of our proposed \mqspn method.
	}
	\vspace{-1em}
	\label{fig:MultiTab}
\end{figure}

We introduce \mqspn, a multi-table \cardest method based on \sys, as illustrated in Figure~\ref{fig:MultiTab}. 
For ease of presentation, this paper considers a query \( q \) that joins two tables \( S \) and \( T \) on the inner join condition \( S.sid = T.tid \), denoted as \( q(S \bowtie T) \), with base table filter predicates \( q(S) \) and \( q(T) \). In particular, we assume that both \( S.sid \) and \( T.tid \) share the same value domain \( D \). Based on this notation, the problem of multi-table \cardest can be formalized as estimating the cardinality \( |q(S \bowtie T)| \), which can be derived as:
\begin{equation} \label{eq:multi-origin}
|q(S \bowtie T)| = |S|  |T| \sum_{v \in D} \Prob(sid = v \wedge q(S)) \cdot \Prob(tid = v \wedge q(T)),
\end{equation}
where \( \Prob(sid = v \wedge q(S)) \) (or \( \Prob(sid = v \wedge q(T)) \)) denotes the probability that the join key \( S.sid \) (or \( T.tid \)) equals \( v \) in the result table of the base filter predicates \( q(S) \) (or \( q(T) \)).

Directly estimating the cardinality \( |q(S \bowtie T)| \) using Equation~(\ref{eq:multi-origin}) is computationally expensive. To address this, \mqspn supports multi-table \cardest by \emph{binning join keys}. Specifically, we divide the domain of the join keys, namely \( sid \) and \( tid \), into a set of \textbf{bins}, denoted as \( \mathcal{B} = \{B_1, B_2, \ldots, B_n\} \). We then estimate \( |q(S \bowtie T)| \) using these bins, \ie
\begin{equation} \nonumber
|q(S \bowtie T)| = |S|  |T| \sum_{B \in \mathcal{B}} \sum_{v \in B} \left\{ \Prob(sid = v \wedge q(S)) \cdot \Prob(tid = v \wedge q(T)) \right\}.
\end{equation}

The task is to estimate \( \sum_{v \in B}  \Prob(sid = v \wedge q(S)) \cdot \Prob(tid = v \wedge q(T)) \) for each individual bin \( B \in \mathcal{B} \).
To achieve this, we propose maintaining basic statistics for each bin \( B \) of values. Formally, we define a bin for a join key, say \( S.sid \), as a triple \( B = (\texttt{id}, \texttt{num}, \texttt{mcv}) \), where \( \texttt{id} \) is the identifier of the bin \( B \), \( \texttt{num} \) is the number of tuples in \( B \), and \( \texttt{mcv} \) is the most common value in \( B \) along with its frequency.
%
 Figure~\ref{fig:MultiTab_b} provides an example of range-based binning: the bin \( B^{S} \) for \( sid \), corresponding to range \([10, 19]\), has \( \texttt{num} = 67 \) and \( \texttt{mcv} = 15 \) with frequency 42. Similarly, we can compute the corresponding bin \( B^{T} \) for the query result \( q(T) \) over table $T$.
%
Then, we can estimate \( \sum_{v \in B_2} \left\{ \Prob(sid = v \wedge q(S)) \cdot \Prob(tid = v \wedge q(T)) \right\} \) based on the two bins \( B^{S}_2 \) and \( B^{T}_2 \).

In this section, we address two challenges in the above estimation process. 
First, while it is straightforward to compute statistics for a given bin \( B \) over a join key, such as \( sid \), the task becomes more complex when considering the base table predicates, such as \( q(S) \), because these predicates may have intricate correlations with the join keys.
%
The second challenge lies in estimating \( \sum_{v \in B} \left\{ \Prob(sid = v \wedge q(S)) \cdot \Prob(tid = v \wedge q(T)) \right\} \) based on the generated bins from \( q(S) \) and \( q(T) \) respectively.

\stitle{Binning Generation.}
To tackle the first challenge, we introduce a \emph{binning generation} method based on our single-table \sys model.
Specifically, given base table predicates, such as \( q(S) \) over table \( S \), Binning Generation generates statistics for each bin \( B \in \mathcal{B} \) corresponding to the data table that satisfies \( q(S) \). To account for the intricate correlations between the predicate \( q(S) \) and the join key, such as \( sid \), this paper proposes a bottom-up traversal of the constructed \sys of table \( S \), as illustrated in Figure~\ref{fig:MultiTab_a}.
Specifically, during the traversal, when a particular node \( \node \) is visited, the key task is to generate a set of bins \( \mathcal{B}_\node \) based on the bins of \( \node \)'s child nodes, and then return \( \mathcal{B}_\node \) to \( \node \)'s parent node. To this end, we propose bin generation strategies tailored to different node types.

(1) If \( \node \) is a \leafnode corresponding to the join key, say \( sid \), we can directly return the pre-generated bins \( \mathcal{B}_\node \) for the join key, as shown in Figure~\ref{fig:MultiTab_a}. Specifically, if query \( q(S) \) over table \( S \) includes predicates on \( sid \), we filter the bins to retain only those that satisfy the predicates.

(2) If \( \node \) is a \prodnode or \qpnode node, at most one child node of \( \node \), say \( \node^\prime \), will return the bins \( \mathcal{B}_{\node^\prime} \) corresponding to its subtree, while the other child nodes return the estimated probabilities \( \Prob_i \) based on the predicates in \( q(S) \). In this case, we scale the \( \texttt{num} \) and \( \texttt{mcv} \) of each bin \( B \in \mathcal{B}_{\node^\prime} \) by a factor of \( \prod_{i}{\Prob_i} \), as shown in Figure~\ref{fig:MultiTab_a}. This scaling is reasonable because the child nodes of \( \node \) are either independent in terms of data correlation or are infrequently co-accessed by the queries.

(3) If \( \node \) is a \sumnode node and its column set \( \Aset_{\node} \) contains the join key, say \( sid \), then all the child nodes of \( \node \) return their respective bins. In this case, we merge the bins from these child nodes as shown in Figure~\ref{fig:MultiTab_b}. Specifically, if bins from different child nodes correspond to the same value range, e.g., \( [10, 19] \), we sum the \( \texttt{num} \) values of these bins to derive a new \( \texttt{num} \). For the \( \texttt{mcv} \), we compute the frequencies of all possible most common values and select the one with the highest frequency as the new \( \texttt{mcv} \).

(4) If \( \node \) is a \qsnode node, we can simply return the bins from its child node to which query \( q(A) \) is routed.

\etitle{Remarks.}
Note that in this section, we focus on the case of a single inner join condition, i.e., \( S.sid = T.tid \). The binning generation method described above can be easily extended to handle multiple inner join conditions between tables \( S \) and \( T \).

\stitle{Multi-Table \cardest based on Binning.} 
To address the second challenge, we propose an effective method to estimate \( \sum_{v \in B} \left\{ \Prob(sid = v \wedge q(S)) \cdot \Prob(tid = v \wedge q(T)) \right\} \) for each ``matched'' bin \( B \), which is shared by both table \( S \) with predicates \( q(S) \) and table \( T \) with predicates \( q(T) \). For ease of presentation, we use $B^{S}$ and $B^{T}$ to denote the matched bins, \ie bins with the same range of join key values. Specifically, we consider the following cases. 

(1) The first case occurs when \( B^{S} \) and \( B^{T} \) share the same \(\texttt{mcv}\), and the \(\texttt{mcv}\) is significant compared to \(\texttt{num}\), \eg the bins with range \([10,19]\) in Figure~\ref{fig:MultiTab_b}. In this case, we can use the \(\texttt{mcv}\) as the representative for the bins, disregarding other values due to their insignificant presence. Based on this, we multiply the frequencies of the \(\texttt{mcv}\) to estimate the result, \eg \( (42 \times 27) / (|S| \times |T|) \).

(2) The second case occurs when \( B^{S} \) and \( B^{T} \) share the same \(\texttt{mcv}\), and the \(\texttt{mcv}\) is significant, \eg the bins with range \([20,29]\). In this case, we cannot directly use the frequencies of the \(\texttt{mcv}\) for estimation because the value distributions of \( B^{S} \) and \( B^{T} \) may differ. To address this, we introduce a scaling factor, \( \min\left( \frac{B^{S}.\texttt{num}}{B^{S}.\texttt{mcv}}, \frac{B^{T}.\texttt{num}}{B^{T}.\texttt{mcv}} \right) \). For example, consider the bins with range \([20,29]\); we scale the multiplying results of the \(\texttt{mcv}\) frequencies by a factor of \( 52/7 \).

(3) The third case occurs when \( B^{S} \) and \( B^{T} \) have different \(\texttt{mcv}\) values, as seen in the bins with the range \([30,39]\) in Figure~\ref{fig:MultiTab_b}. In this case, we focus on the \(\texttt{mcv}\) with the larger frequency and assume that the tuple count of the other values in the bin comes from the non-\(\texttt{mcv}\) values. Therefore, the tuple count for this match is $\frac{B^{S}.\texttt{num}-B^{S}.\texttt{mcv}}{|B^{S}.\texttt{range}| - 1} \times B^{T}.\texttt{mcv}$. This applies when \(B^{T}.\texttt{mcv} > B^{S}.\texttt{mcv}\).

%% file: secs/expres.tex
\section{Experiments}\label{sec:exp}

\subsection{Experimental Setting}



\begin{table}
  \caption{Statistics of Datasets.}
  \label{tab:Expr_datasets}
  \vspace{-1em}
  \resizebox{0.98\columnwidth}{!}{
  \begin{tabular}{|c||c|c|c|c|}
    \hline
    \textbf{Dataset} & \textbf{Tables} & \textbf{Tuples} & \textbf{Columns (Numeric)} & \textbf{Domain Size} \\
    \hline \hline
    GAS & 1 & $4$M & $8$ $(8)$ & $10^1\sim10^2$ \\ 
    \hline
    Census & 1 & $0.05$M & $14$ $(5)$ & $10^1\sim10^4$ \\ 
    \hline
    Forest & 1 & $0.6$M & $54$ $(10)$ & $10^2\sim10^3$ \\ 
    \hline
    Power & 1 & $2$M & $9$ $(7)$ & $10^4\sim10^5$ \\ 
    \hline
    IMDB & 21 & $74$M & $108$ $(57)$ & $10^0\sim10^7$ \\ 
  \hline
\end{tabular}
\vspace{-2em}
}
\end{table}




\stitle{Datasets.} We evaluate our proposed \sys approach on both single-table datasets and multi-table datasets. Table~\ref{tab:Expr_datasets} provides the statistics of the datasets used in our experiments. 
%

\etitle{Single-table Datasets.} We use the following four single-table datasets.
(1) \emph{GAS}~\cite{gas} is a real-world gas sensing dataset, and we extract the most informative 8 columns (Time, Humidity, Temperature, Flow\_rate, Heater\_voltage, R1, R5, and R7), following the existing works~\cite{flat,flat-code}.
(2) \emph{Census}~\cite{census} is a dataset about income, extracted by Barry Becker from the real-world 1994 Census database.
(3) \emph{Forest}~\cite{forest} is a real-world forest-fire dataset from the US Forest Service (USFS) and US Geological Survey (USGS).
(4) \emph{Power}~\cite{power} is an electric power consumption dataset consisting of measurements gathered in a house located in Sceaux (7 km from Paris, France) between December 2006 and November 2010.

%

\etitle{Multi-table Dataset.} We evaluate multi-table cardinality estimation using the \emph{IMDB}~\cite{imdb} dataset, a real-world dataset containing 50K movie reviews. This dataset is extensively used in existing works~\cite{postgres,deepdb,flat,mscn,neurocard,uae} for multi-table cardinality estimation evaluation. The columns in \emph{IMDB} typically have large domain sizes, which presents a greater challenge for \cardest.

\stitle{Query Workloads.} 
We describe below how the query workloads used in the experiments are prepared.

\etitle{Synthetic Workloads for Single-Table \cardest.}
As there is no real query workload available for the above four single-table datasets, we use the following steps to synthesize workloads.

(1) {Template generation}: Following existing DBMS benchmarks~\cite{tpch,tpcds,joblight}, we first generate SQL templates containing different column combinations as query predicates, and then synthesize queries based on these templates.

(2) {Template selection}: Existing studies~\cite{mscn,naru,deepdb,arewe-paper,flat} have shown that data correlation significantly affects the performance of \cardest methods.
To account for this, we cluster all generated templates into two groups: one containing templates that access highly correlated columns and the other containing templates that access weakly correlated columns. We then evenly sample templates from both groups to ensure a fair comparison.

(3) {Query synthesis}: 
Given a template, we generate queries by filling it with randomly selected predicates. Previous studies~\cite{arewe-paper,naru,uae} have shown that query conditions significantly affect \cardest performance. To ensure a fair comparison, we follow the method from~\cite{arewe-paper}, generating query conditions in two steps: first, selecting a random tuple from the data table as the center of the range; then, determining the width of each range using either a uniform or exponential distribution, with a predefined ratio controlling the selection between them. This ensures the conditions cover a variety of query patterns for evaluation.


We generate a read-write hybrid workload to evaluate \sys model updates, where each SQL statement is randomly selected to be either a query or a DML command (INSERT or DELETE). First, we create a training set with queries accessing weakly correlated columns and range constraints following a normal distribution. Then, we add at least 20\% new data tuples with increased correlation by sampling from a table with sorted columns. These new tuples are included as DML commands. Finally, we generate a test set that includes (1) queries following the original patterns and distribution and (2) queries accessing highly correlated columns with range constraints following an exponential distribution.

\etitle{Real-World Workload for Multi-Table \cardest.}
For our multi-table dataset \emph{IMDB}, we use \emph{JOB-light}~\cite{joblight}, the most widely adopted multi-table workload for \cardest. \emph{JOB-light} consists of 70 queries on the \emph{IMDB} dataset, where each query includes joins on 2 to 5 tables, along with 1 to 5 range constraints. These queries present a significant challenge for cardinality estimation methods. In particular, to ensure a fair comparison with the hybrid-driven model \emph{UAE}, we use \emph{JOB-light} as the workload test set and adopt the extended workload provided by \emph{UAE} as the workload training set. This extended workload contains 100,000 queries covering \emph{JOB-light}.

%

\stitle{Baselines.} 
We compare \sys against the following baselines.

\etitle{MSCN~\cite{mscn}} is a query-driven \cardest model based on regression models.  
We use the implementation of MSCN from~\cite{arewe-paper} and set its hyper-parameters following the configurations in~\cite{arewe-paper}.

\etitle{Naru~\cite{naru}} is a DAR-based model that fits the joint data distribution to compute \cardest.  
We use the implementation of Naru from~\cite{arewe-paper} and set its hyper-parameters following the configurations in~\cite{arewe-paper}.

\etitle{DeepDB~\cite{deepdb}} is a data-driven SPN-based model which fits joint data distribution to compute \cardest, following local independence assumption. We use the implemention of DeepDB from~\cite{arewe-paper} and set its hyper-parameters following the configurations in~\cite{arewe-paper}.

\etitle{FLAT~\cite{flat}} is a \cardest method based on DeepDB, which introduces factorization and multi-dimensional histograms.
We use the implementation provided by the authors of FLAT~\cite{flat-code} and tune its hyper-parameters as recommended in the original paper.

\etitle{UAE~\cite{uae}} is a hybrid-driven \cardest method based on Naru, combining unsupervised losses from data with supervised losses from queries.  
We use the implementation provided by the authors of UAE~\cite{uae-code}
We also use it as multi-table \cardest method~\cite{uaejoins}.

\etitle{FactorJoin~\cite{factorjoin}} is a multi-table \cardest method based on join-keys binning without modeling data distribution on generated outer-join tables. Following the paper~\cite{factorjoin}, we implement it with uniform sampling (sample\_rate=0.1).

\etitle{Postgres~\cite{postgres}} is the cardinality estimator of PostgreSQL, based on traditional statistics-based methods.  
We run and evaluate it using the connector from~\cite{arewe-paper}, which connects to PostgreSQL 12 with the default setting \( stat\_target=10000 \). 

\etitle{Sampling} is a traditional \cardest method based on random sampling of data.  
We use the implementation of Sampling from~\cite{arewe-paper} with the default setting \( ratio=0.015 \).  

\etitle{MHist} is a traditional \cardest method that uses a single multi-dimensional histogram to model the joint data distribution.  
We use the implementation of MHist from~\cite{arewe-paper} with the default setting. 

\stitle{Evaluation Metrics.} We use the following metrics to comprehensively evaluate and compare \cardest methods.
%

\etitle{Estimation Accuracy.}
For each query, we measure estimation accuracy using Q-error, defined as the ratio between the estimation \( est \) and the ground truth \( gt \), \ie $Q\text{-error} = \frac{\min\{est, gt\}}{\max\{est, gt\}}$.

%


\etitle{Inference Time.} For each query in the test set, we measure the total time spent for \cardest as inference time. We then compute the mean inference time among the queries in the test set.


\etitle{Storage Overhead.} We measure the total size of the model or statistics file(s) for each method on each dataset and its corresponding workload training set, defining this as the storage overhead.





\stitle{Experimental Settings.} 
All evaluated methods are implemented in Python 3.7. Our experiments are conducted on an AMD-64 server with the following specifications:  
OS: Ubuntu 20.04.6 LTS;  
Dual CPU System: \(2 \times\) Intel(R) Xeon(R) Gold 6230 CPU @ 2.10GHz (20C40T);  
Main Memory: 1TB DDR4 ECC;  
Storage: \(4 \times 8\) TB HDD (RAID5).  
We set the default hyper-parameters of \sys (query-aware adaptive threshold of \prodnode RDC, threshold of \qpnode, threshold of \qsnode, and threshold of \sumnode) on all datasets as: $\tau_{p} = (s=5, l=0.1, u=0.3), \tau = 0.01, \tau_{x} = 0.7, \tau_{s} = 0.3$.

%

\subsection{Evaluation on Single-Table \cardest}

\input{table/main_exp_template}

\input{table/update_exp}



We first compare \sys with the baseline methods on single-table \cardest.  
Table~\ref{tab:single_all} reports the experimental results.




\stitle{Estimation Accuracy.}
Our proposed \sys achieves superior estimation accuracy, comparable to state-of-the-art (SOTA) data-driven methods (\emph{FLAT} and \emph{Naru}) and the hybrid method \emph{UAE}. Specifically, the mean Q-errors of \sys are minimal, ranging from 1.0 to 1.8. Moreover, \sys overcomes the limitations of traditional SPN models (e.g., \emph{DeepDB}), achieving up to an 88\% reduction in Q-error. The superior performance of \sys is primarily attributed to its ability to effectively handle strongly correlated data through column partitioning based on both data distribution and query co-access patterns. 
In contrast, query-driven methods such as \emph{MSCN} and traditional \cardest methods (Postgres, Sampling, and MHist) struggle to achieve satisfactory estimation accuracy in most cases, as they fail to adequately learn the data distribution. Additionally, we observe that the Q-errors of \emph{DeepDB} on the \emph{GAS} and \emph{Power} datasets rise sharply. This is due to the strong correlations present in these datasets, which pose a more significant challenge for \emph{DeepDB}.

\stitle{Inference Time.}
As shown in Table~\ref{tab:single_all}, \sys demonstrates highly efficient inference performance. Specifically, the inference time of \sys ranges from 0.422 ms to 3.092 ms across the four datasets, which is only slightly slower than the traditional \cardest method \emph{Postgres} and the query-driven method \emph{MSCN}. Moreover, compared to the traditional SPN-based method \emph{DeepDB}, \sys achieves up to a 92.2\% reduction in inference time. 
This efficiency is due to \sys's column partitioning strategy, which considers both data correlations and query access patterns. By reducing the number of intermediate nodes in the SPN, \sys improves inference efficiency and reduces storage overhead, all while maintaining high estimation accuracy. 
On the other hand, \emph{Naru} and \emph{UAE} are the slowest methods, as their underlying deep auto-regressive models suffer from high inference times due to the computationally expensive progressive sampling process. Additionally, the inference time of \emph{FLAT} increases significantly on the \emph{Forest} and \emph{Power} datasets, due to the large number of factorize and multi-leaf nodes used to model the complex correlations in these datasets.

We also evaluate the construction (training) time of the methods and categorize them into three groups:  
(1) \emph{Postgres}, \emph{Sampling}, and \emph{MSCN} require negligible time for training.  
(2) SPN-based methods, such as \emph{DeepDB}, \emph{FLAT}, and \sys, take approximately 1-3 minutes for construction, which does not impose a significant burden on the DBMS.  
(3) DAR-based methods like \emph{Naru} and \emph{UAE} require 100 to 1000 times more training time than SPN-based models, making them impractical for real-world applications. 

\stitle{Storage Overhead.} 
\sys ranks among the top in storage efficiency, requiring only tens of KB to about 1 MB more than \emph{Postgres}. \emph{DeepDB} can also be considered a lightweight method, although its model size increases significantly on the \emph{Forest} and \emph{Power} datasets due to a larger number of nodes. 
%
%
On the other hand, \emph{MSCN}, \emph{Naru}, and \emph{UAE}, which are based on CNN or DAR models, naturally have larger model sizes. \emph{FLAT} and \emph{Sampling} suffer from substantial storage overhead. The excessive model size of \emph{FLAT} stems from the same factors that contribute to its slow performance. 

\stitle{Summary.} The experimental results demonstrate that \sys achieves superior and robust performance across the three key criteria, outperforming state-of-the-art approaches. This outcome aligns with the design objectives of \sys, as presented in Table~\ref{tab:intro_comp}.

\input{table/join_exp}

\subsection{Evaluation on Dynamic Model Update}


In this section, we evaluate the model update performance of \sys on our read-write hybrid workload and compare the following alternatives: \emph{NoTrain} (no model updates), \emph{ReTrain} (periodic model reconstruction), and \emph{AdaIncr} (our updating method in \sys).
Due to space constraints, we only report the results for the hybrid-update setting, which involves both data updates and query workload shifts, while the results for the data-update and query-update settings are provided in our technical report.
As shown in Figure~\ref{fig:update_qerr_dq}, \emph{NoTrain} suffers from poor accuracy across all four datasets under hybrid workloads, whereas both \emph{ReTrain} and \emph{AdaIncr} maintain the accuracy of the \sys model.  
Notably, \emph{AdaIncr} reduces update time by 30\% to 60\% while achieving the same accuracy as \emph{ReTrain}. Furthermore, when analyzing the trends in workload execution time, the total time usage of \emph{AdaIncr} increases gradually, whereas that of \emph{ReTrain} rises sharply.

\subsection{Evaluation of Multi-Table \cardest}


In this section, we compare our proposed \mqspn with the SOTA methods for multi-table \cardest and report the results in Table~\ref{tab:multi_all}. 
We observe that \emph{Postgres} exhibits poor estimation accuracy, and loses its advantage in inference time, as it requires multiple iterations to compute multi-table join cardinality.  
\emph{UAE} achieves better estimation accuracy; however, both its inference and training times are unsatisfactory due to the inherent limitations of DAR-based models. 
Compared to \emph{UAE}, \emph{FLAT} provides a better trade-off between estimation accuracy and inference time.
%
%
Among all evaluated approaches, our proposed \mqspn model achieves the best performance. Specifically, compared to the best baseline, \emph{FLAT}, \mqspn improves estimation accuracy and inference efficiency by approximately three times, covering the shortage of \emph{FactorJoin} on base table filtering. The only drawback of \mqspn is its model size, primarily due to storing binnings on \leafnode nodes with extra storage for convenience. However, this additional storage overhead can be eliminated by integrating binning into the histogram on each \leafnode node during implementation.

\subsection{Evaluation on End-to-End Query Execution} 

\begin{figure}[t!]
	\centering
	\includegraphics[width=0.9\columnwidth]{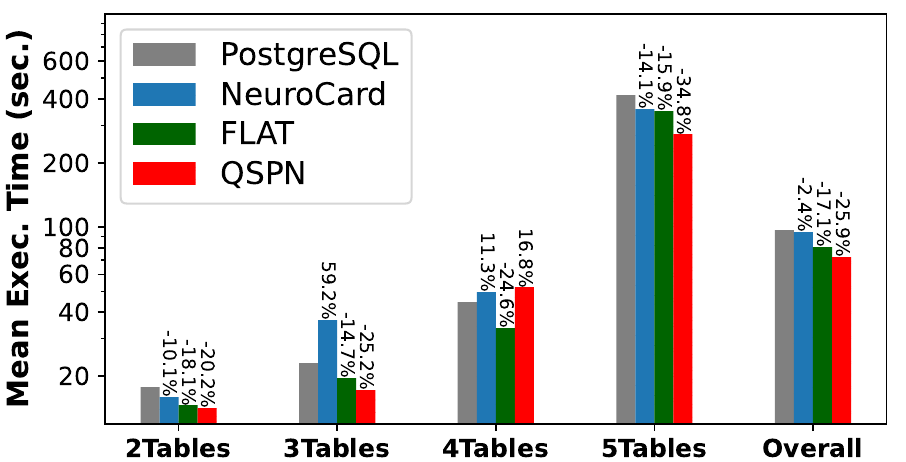}
	\vspace{-1em}
	\caption{End-to-End Mean Query Execution Time.}
	\vspace{-2em}
	\label{fig:expres_end2end}
\end{figure}

We evaluate the effect of \cardest models on end-to-end query execution in the PostgreSQL DBMS~\cite{postgres}, following the experimental settings described in the End-to-End \cardest Benchmark~\cite{end2end}. 
Specifically, we conduct the End-to-End benchmark on the \emph{IMDB} dataset~\cite{imdb} with \emph{JOB-Light} workload~\cite{joblight}, and compare our proposed \sys with the PostgreSQL internal estimator~\cite{postgres}, \emph{NeuroCard}~\cite{neurocard}, and \emph{FLAT}~\cite{flat} in the End-to-End evaluation. 
The experimental results are reported in Figure~\ref{fig:expres_end2end}. Our \sys achieves the best performance in end-to-end query execution. Specifically, \sys reduces the mean query execution time by $25.9\%$, compared to $2.4\%$ for \emph{Neurocard} and $17.1\%$ for \emph{FLAT}. This outcome aligns with the superior performance of our \sys in multi-table \cardest.
In particular, \sys performs best on queries involving 2, 3, and 5 joined tables. For instance, for the most complex 5-table queries, our \sys reduces execution time by more than $15\%$ compared to \emph{FLAT}. 

%% file: table/main_exp_template.tex
\begin{table*}[t]
	\caption{Evaluating Single-Table \cardest on Key Criteria: Estimation Errors, Inference/Construction Time and Model Size.}
	\label{tab:single_all}
	\vspace{-1em}
         \resizebox{0.85\textwidth}{!}{
         \renewcommand{\arraystretch}{0.9} 
         \small
	\begin{tabular}{|c|c||cccccc||c|c||c|}
    \hline
\multirow{2}*{\textbf{~~Dataset~~}}  &  \multirow{2}*{\textbf{~~~~Method~~~~}}  &  \multicolumn{6}{c|}{\textbf{Estimation Errors}}  &  \textbf{~~Inference~~}  &  \textbf{~~Construction~~}  &  \textbf{~~Model~~} \\  \cline{3-8} 
  &    &  \textbf{50th}  &  \textbf{90th}  &  \textbf{95th}  &  \textbf{99th}  &  \textbf{Max}  &  \textbf{Mean}  &  \textbf{Time (ms)}  &  \textbf{Time (min.)}  &  \textbf{Size (MB)} \\
 \hline
 \hline 
\multirow{9}*{GAS}  &  Postgres  & 3.42 & 33.37 & 48.46 & 143.34 & 376.21 & 13.96 & \underline{0.321} &  0.123  & \textbf{0.01} \\
  &  Sampling  & \textbf{1.02} & \textbf{1.14} & \textbf{1.19} & \textbf{1.54} & \textbf{1.97} & \textbf{1.06} & 1.347 &  \textbf{0.002}  & 250.118 \\
  &  MHist  & 1.84 & 4.63 & 12.97 & 290.38 & 5564 & 28.21 & 187.55 &  17.600  & 3.09 \\
 \cline{2-11}
  &  MSCN  & 1.46 & 3.61 & 6.45 & 22.21 & 43.96 & 2.65 &  0.386  &  0.157  & 4.423 \\
  & LW-XGB & 1.46 & 3.23 & 4.27 & 12.38 & 154.22 & 2.81 & \textbf{0.066} & \underline{0.08} & 0.605 \\
 \cline{2-11}
  &  Naru  & 1.07 &  \underline{1.26}  &  1.39  &  1.87  &  2.47  &  1.12  &  13.658  &  149.205  & 1.923 \\
  &  DeepDB  &  \underline{1.06}  & 9.32 & 14.22 & 48.23 & 83.05 & 3.94 & 8.22 &  3.316  &  0.580 \\
  &  FLAT  &  \textbf{1.02}  & 1.7 & 2.12 & 3.12 & 3.94 & 1.21 & 1.209 &  0.810  & 13.996 \\
 \cline{2-11}
  &  UAE  &  \underline{1.06}  &  \underline{1.21}  &  \underline{1.30}  &  \underline{1.71}  &  \underline{2.14}  &  \underline{1.10}  & 2.013 &  301.390  & 0.747 \\
  &  QSPN  &  \underline{1.06}  & 1.69 & 2.12 & 3.12 & 3.94 & 1.22 &  1.089  &  1.919  &  \underline{0.209}  \\
\hline \hline
\multirow{9}*{Census}  &  Postgres  & 6.75 & 118.1 & 393.67 & 2362 & 2362 & 91.25 & \underline{0.133} &  0.045  & \textbf{0.01} \\
  &  Sampling  & 1.18 & 6.31 & 24.6 & 72.58 & 97 & 4.71 & 0.893 &  \textbf{0.001}  & 1.994 \\
  &  MHist  & \textbf{1} & \textbf{1} & \textbf{1} & \textbf{1} & \textbf{1} & \textbf{1} & 68.69 &  4.830  & 0.98 \\
 \cline{2-11}
  &  MSCN  & 2.37 & 6.25 & 13.07 & 78 & 79 & 4.85 &  0.391  &  0.108  & 0.085 \\
 & LW-XGB & 1.05 & 2.62 & 4.33 & 11.07 & 13.23 & 1.65 & \textbf{0.053} & \underline{0.01} & 0.138\\
 \cline{2-11}
  &  Naru  &  \underline{1.03}  &  1.23  &  1.40  &  \underline{1.69}  &  3.00  &  \underline{1.09}  & 5.985 &  1.522  & 0.128 \\
  &  DeepDB  & 1.12 & 1.54 & 1.78 & 3 & 6.86 & 1.25 & 0.438 &  0.090  &  \underline{0.026} \\
  &  FLAT  & 1.11 & 1.4 & 1.72 & 3 & 6.67 &  1.23  & 0.938 &  0.107  & 0.104 \\
 \cline{2-11}
  &  UAE  &  1.06  &  \underline{1.15}  &  \underline{1.25}  &  1.76  &  \underline{2.67}  &  \underline{1.09}  & 2.28 &  13.044  & 0.829 \\
  &  QSPN  & 1.12 & 1.42 & 1.7 & 3 & 6.67 &  1.23  &  0.422  &  0.141  &  0.059 \\
\hline \hline
\multirow{9}*{Forest}  &  Postgres  & 3.43 & 40.67 & 109.74 & 1002.98 & 9215.5 & 64.72 & \underline{0.153} &  0.158  & \textbf{0.07} \\
  &  ~~~~~Sampling~~~~~  & 1.07 & \textbf{1.44} & 1.89 & 25.01 & 169 & 2.64 & 0.994 &  \textbf{0.001}  & 47.378 \\
  &  MHist  & 1.59 & 5.46 & 10.12 & 40.87 & 1427 & 8.86 & 420.915 &  20.000  & 3.83 \\
 \cline{2-11}
  &  MSCN  & 1.49 & 3.53 & 5.36 & 11.73 & 61.5 & 2.25 &  0.427  &  0.128  &  1.501 \\
 & LW-XGB & 1.44 & 4.63 & 11.03 & 41.92 & 139.43 & 3.63 & \textbf{0.074} & \underline{0.01} & \underline{0.663}\\
 \cline{2-11}
  &  Naru  & 1.17 & 1.79 & 2.31 & 7.39 &  \textbf{22.00}  &  \underline{1.50}  & 38.651 &  80.026  & 2.216 \\
  &  DeepDB  & 1.07 &  \underline{1.57}  & 2.02 & 4.81 & 148 & 1.88 & 11.295 &  2.562  & 1.64 \\
  &  FLAT  &  \textbf{1.02}  &  \underline{1.48}  &  \textbf{1.80}  &  \underline{3.67}  & 38.71 &  \textbf{1.30}  & 41.69 &  1.821  & 972.307 \\
 \cline{2-11}
  &  UAE  & 1.16 & 1.82 & 2.75 & 12.96 &  \underline{25.20}  & 1.6 & 76.721 &  188.661  & 29.147 \\
  &  QSPN  &  \underline{1.05}  &  \underline{1.48}  &  \underline{1.84}  &  \textbf{3.08}  & 186.5 & 1.81 &  3.092  &  2.167  &  1.261 \\
\hline \hline
\multirow{9}*{Power}  &  Postgres  & 3.71 & 162.04 & 925.33 & 13374.34 & 48341 & 551.04 & \textbf{0.103} &  0.206  & \textbf{0.04} \\
  &  Sampling  & \underline{1.04} & 1.42 & 4 & 60.96 & 394 & 3.96 & 1.147 &  \textbf{0.001}  & 118.273 \\
  &  MHist  & 15 & 104.73 & 349.63 & 5350.84 & 15829 & 211.63 & 182.491 &  13.350  & 2.72 \\
\cline{2-11}
  &  MSCN  & 1.88 & 8.31 & 15.58 & 32.45 & 155 & 4.32 &  \underline{0.421}  &  0.144  & 4.402 \\
   & LW-XGB & 1.34 & 5.59 & 14.14 & 46.49 & 242.08 & 4.204 & 0.466 & \underline{0.02} & \underline{0.062}\\
\cline{2-11}
  &  Naru  & 1.06 & 1.35 & 1.61 &  \underline{3.48}  &  \underline{7.95}  &  \underline{1.19}  & 43.125 &  175.368  & 4.327 \\
  &  DeepDB  &  \underline{1.04}  & 3.42 & 5.45 & 198.2 & 1876 & 13.05 & 12.731 &  3.722  &  \underline{3.711} \\
  &  FLAT  &  \textbf{1.01}  &  \textbf{1.22}  &  \textbf{1.39}  &  \textbf{2.27}  & 50.59 & 1.27 & 2090.254 &  2.225  & 10100.46 \\
\cline{2-11}
  &  UAE  &  \underline{1.04}  &  \underline{1.27}  &  \underline{1.58}  & 4.15 &  \textbf{5.00}  &  \textbf{1.16}  & 18.22 &  1316.239  & 12.599 \\
  &  QSPN  & 1.05 & 1.74 & 2.57 & 7.98 & 51.33 & 1.51 &  0.988  &  0.782  &  0.623 \\
  \hline
	\end{tabular}
 }
\end{table*}

%% file: table/update_exp.tex
\begin{figure*}
    \centering
    \includegraphics[width=0.95\linewidth]{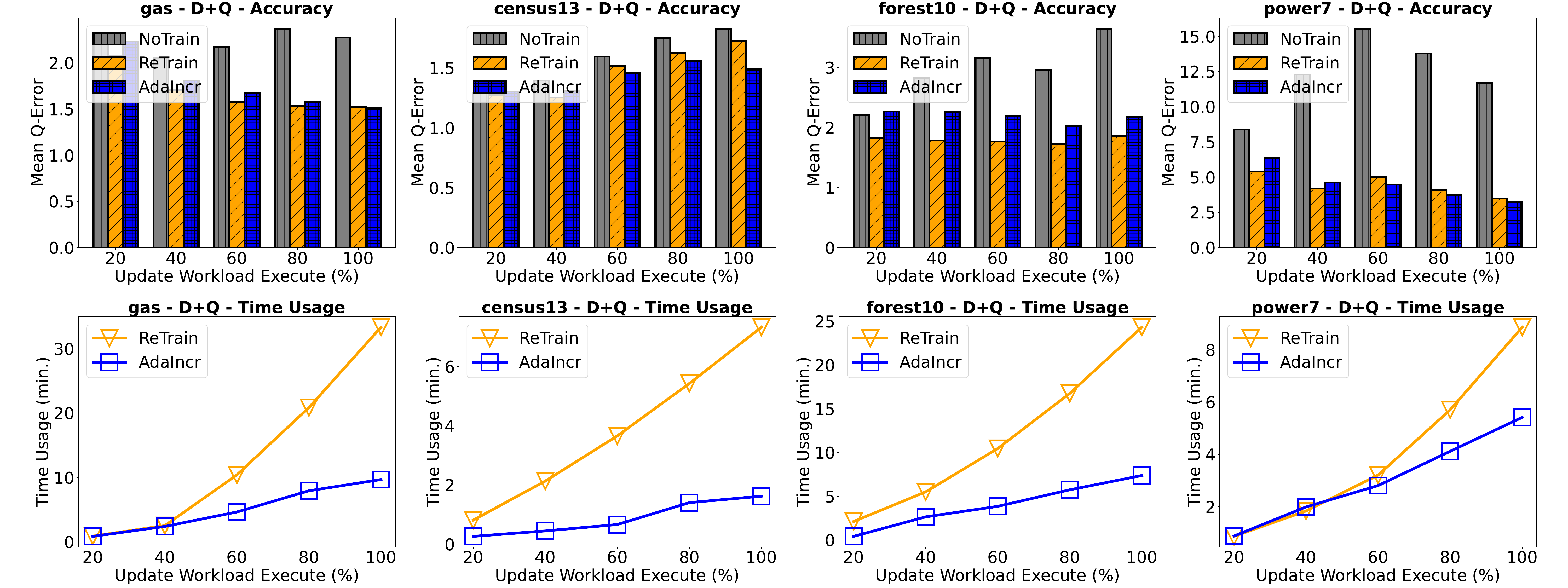}
    \caption{Evaluation on Dynamic Model Update (Hybrid-Update Setting).}
    \label{fig:update_qerr_dq}
\end{figure*}

%% file: table/join_exp.tex
\begin{table*}[t]
	\caption{Evaluating Multi-Table \cardest on the Job-Light Workload of IMDB Dataset.}
	\vspace{-1em}
	\label{tab:multi_all}
         \resizebox{0.85\textwidth}{!}{
        \renewcommand{\arraystretch}{0.9} 
         \small
	\begin{tabular}{|c||cccccc|c|c|c|}
\hline
\multirow{2}*{\textbf{Method}}  &  \multicolumn{6}{c|}{\textbf{Q-error}}  &  \textbf{Inference}  &  \textbf{Construction} &   \textbf{Model} \\ \cline{2-7}
  &  \textbf{50th}  &  \textbf{90th}  &  \textbf{95th}  &  \textbf{99th}  &  \textbf{Max}  &  \textbf{Mean}  &  \textbf{Time (ms)}  &  \textbf{Time (min.)}  &  \textbf{Size (MB)} \\
 \hline
 \hline
 Postgres  & 7.99 & 161.79 & 818.29 & 2042.92 & 2093.13 & 128.61 & 1894.532 &  \underline{3.323}  & \textbf{9} \\
 \hline
 FLAT  & 2.73 & \underline{16.71} & 44.52 & 120.91 & 203.11 & \underline{10.33} & \underline{32.081} &  46  & 91 \\
 FactorJoin  &  4.33  &  24.7  &  34.62  &  \underline{92.57}  &  \underline{106.36}  &  10.63  &  8626.417  &  \textbf{0.426}  &  \underline{13.213} \\
 \hline
 UAE  & \textbf{1.45} & 17.74 & \underline{29} & 184.42 & 184.95 & 11.72 & 123.6 &  27224  &  - \\
 \mqspn  & \underline{2.54} & \textbf{6.5} & \textbf{9.01} & \textbf{26.79} & \textbf{33} & \textbf{3.87} & \textbf{13.368} &  27  & 450 \\
\hline 
	\end{tabular}
 }
\end{table*}

%% file: secs/rw.tex
\section{Related Work} \label{sec:rw}

\stitle{Traditional \cardest Methods.} 
Traditional \cardest methods~\cite{postgres,sample1,sample2,sample3,sample4,mhist,bayes} rely on simplifying assumptions, such as column independence.
Postgres~\cite{postgres} assumes that all columns are independent and estimates the data distribution of each column using histograms.
Sampling-based methods~\cite{sample1,sample2,sample3,sample4} sample tuples from the data and store them. In the online phase, these methods execute queries on the stored sample to estimate cardinalities.
MHist~\cite{mhist} is a multi-histogram approach that accounts for data correlation by constructing multi-dimensional histograms.
Bayes~\cite{bayes} performs cardinality estimation using probabilistic graphical models~\cite{bayes-gm1,bayes-gm2,bayes-gm3}. While effective, it can be slow, especially when dealing with datasets that have high correlation.
The key limitation of these traditional methods is their reliance on simplifying assumptions, which often lead to significant estimation errors.

\stitle{Learning-based \cardest Methods.}
Query-driven methods transform the \cardest problem into a regression task, mapping query workloads to ground truth cardinalities. MSCN~\cite{mscn} encodes a query into a standardized vector, which is then fed into a Multi-Layer Perceptron (MLP), where they undergo average pooling. The pooled representations are then concatenated and passed into a final MLP to output the selectivity.
LW-XGB/NN~\cite{lw} encodes a query into a simpler vector by concatenating the lower bounds and upper bounds of all range predicates in order. Then, this query vector is then input into a small neural network or XGBoost~\cite{xgboost} model to predict the estimated cardinalities.
Data-driven methods transform the \cardest problem into a joint probability problem, where each column is treated as a random variable. Naru~\cite{naru} factorizes the joint distribution into conditional distributions using Deep AutoRegressive (DAR) models such as MADE~\cite{made}. Naru then employs progressive sampling~\cite{progsample} to compute cardinality estimates for range queries based on point probabilities. 
UAE~\cite{uae} is an optimized version of Naru, tailored for query workloads. As a hybrid \cardest model, UAE improves the fitting of the DAR model on data with long-tail distributions by adjusting the sampling regions according to the queries.
FactorJoin~\cite{factorjoin} is a multi-table \cardest method that trains single-table cardinality estimation models for each of the joined tables and uses a factor graph to capture the relationships and correlations between different join keys.
Despite advancements in learning-based cardinality estimation, existing methods may struggle to simultaneously optimize the key criteria: \emph{estimation accuracy}, \emph{inference time}, and \emph{storage overhead}, limiting their practical applicability in real-world database environments.

%% file: secs/conclusion.tex
\section{Conclusion}
In this paper, we have introduced \sys, a \emph{unified model} that integrates both data distribution and query workload. 
\sys extends the simple yet effective Sum-Product Network (SPN) model by jointly partitioning columns based on both data correlations and query access patterns, thereby reducing model size and improving inference efficiency without sacrificing estimation accuracy.
We have formalized \sys as a tree-based structure that extends SPNs by introducing two new node types: \term{QProduct} and \term{QSplit}.
We have conducted extensive experiments to evaluate \sys in both single-table and multi-table cardinality estimation settings. The experimental results have demonstrated that \sys achieves superior and robust performance on the three key criteria, compared with state-of-the-art approaches.





%% file: reference.tex

%% file: secs/appendixA.tex
\section{Additional Update Experiments}

\begin{figure*}
	\centering
	\includegraphics[width=1.0\linewidth]{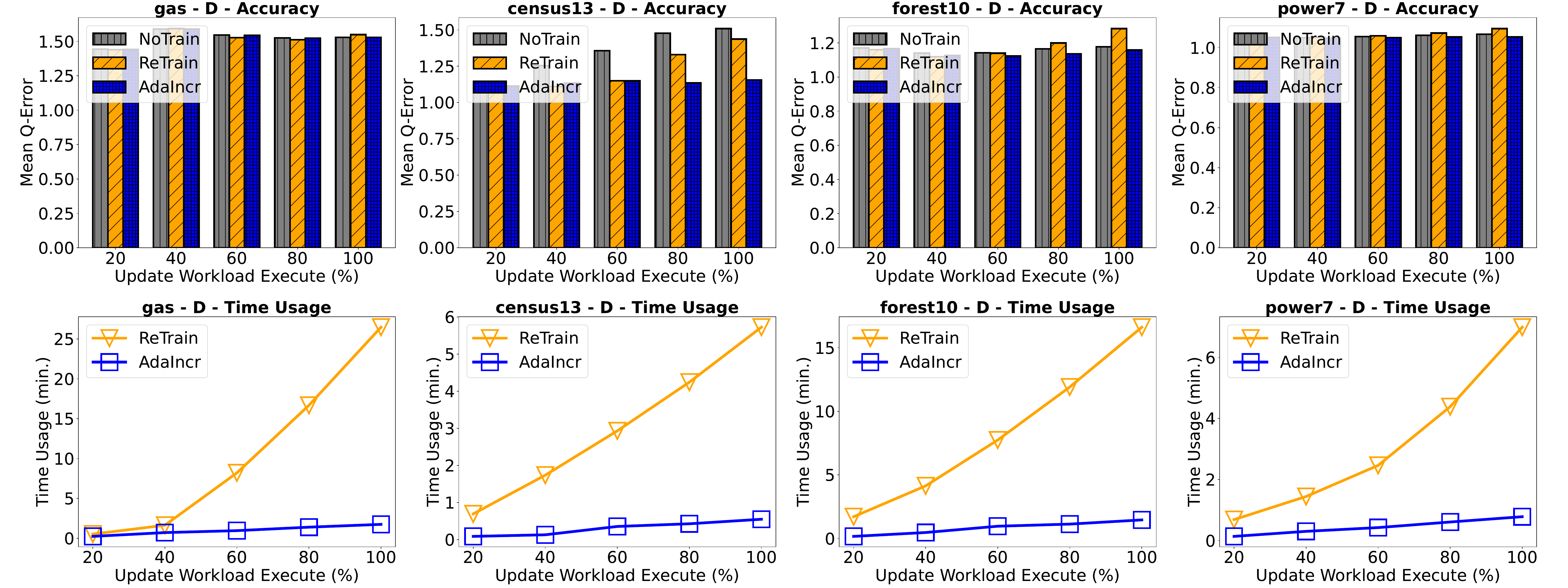}
	\caption{Evaluation on Dynamic Model Update (Data-Update Setting).}
	\label{fig:update_qerr_d}
\end{figure*}

\begin{figure*}
	\centering
	\includegraphics[width=1.0\linewidth]{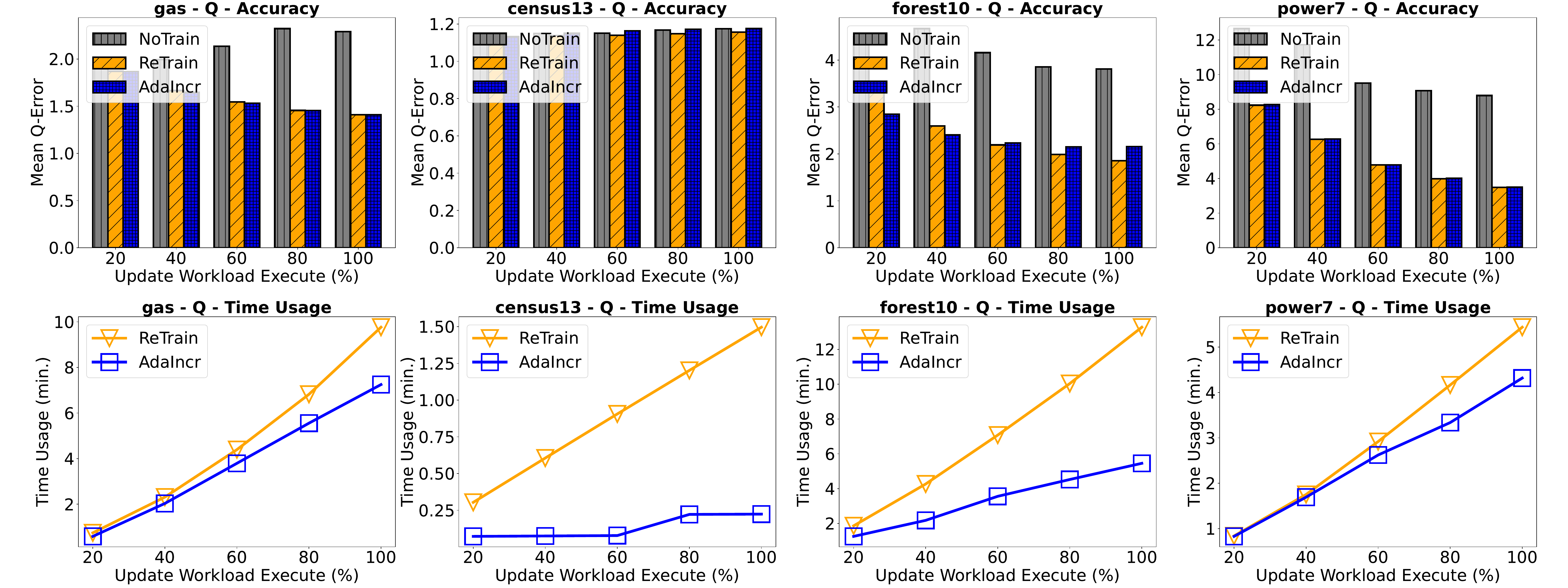}
	\caption{Evaluation on Dynamic Model Update (Query-Update Setting).}
	\label{fig:update_qerr_q}
\end{figure*}

\subsection{Data-Update Only}

We first consider the setting where only data tuples are updated (\ie, \( \Delta T \)).  
As shown in Figure~\ref{fig:update_qerr_d}, data updates have a minimal impact on \sys. This is because, in large datasets with high-cardinality attributes, incremental updates typically do not significantly alter the overall data distribution.
\emph{NoTrain} results in a significant Q-error, whereas both \emph{ReTrain} and \emph{AdaIncr} maintain the accuracy of the \sys model. Compared with \emph{ReTrain}, our \emph{AdaIncr} achieves over a \( 10\times \) reduction in update time while keeping the \sys model even more accurate than \emph{ReTrain}, as its efficiency allows for more frequent updates.




\subsection{Query-Update Only}

As shown in \autoref{fig:update_qerr_q}, query updates do impact \sys because they alter the query-column access patterns. \emph{NoTrain} is not a viable option, as it leads to a sharp increase in mean Q-error. As expected, both \emph{ReTrain} and \emph{AdaIncr} maintain the accuracy of the \sys model, with \emph{AdaIncr} significantly reducing the time required for model updates.  
However, compared to the data-update-only scenario, \emph{AdaIncr} requires more time for model updates in this case, as more subtrees need to be reconstructed to adapt to workload shifts. In contrast to the data-update-only scenario, the \sys model remains unaffected by workload shifts on the \emph{Census} dataset, as it constructs only a few \qpnode and \qsnode nodes due to the weak correlations in the data.

%% file: secs/appendixB.tex
\section{Additional Proof}

\setcounter{lemma}{0}

\subsection{PROOF of LEMMA 1}

\begin{lemma}
	The problem of \qpnode construction is equivalent to the minimum $k$-cut problem. 
\end{lemma}

\begin{proof}
	
	Recall that the minimum $k$-cut problem is defined as follows: Given an undirected graph $G=(V,E)$ with edge weights $w:E \rightarrow \mathbb{R}$ and an integer $k$, the goal is to partition the vertex set $V$ into $k$ disjoint subsets ${C_1, C_2, \ldots, C_k}$ such that the following objective is minimized:
    $
       \sum_{i=1}^{k-1} \sum_{j=i+1}^{k} \sum_{v_1 \in C_i, v_2 \in C_i}{w(\{v_1, v_2\})},
    $
    which corresponds to the sum of weights of edges crossing different partitions.
    
    In the context of \qpnode construction, the column set $\Aset$ and the access affinity $\aff(a_i, a_j)$ between two columns can be naturally mapped to the vertex set $V$ and edge weights $w({v_1, v_2})$ in the minimum $k$-cut problem, respectively. Moreover, the objective of minimizing the inter-partition affinity (IPA) in \qpnode construction is equivalent to the objective of the minimum $k$-cut problem. 
    Thus, we prove the lemma.
    
   We prove that the \qpnode construction problem can be reduced from K-CUT problem. Given an arbitrary undirected graph $G=(V,E)$ and parameter $K$, we make a column set $A=\{a_1,\ldots,a_{|V|}\}$ corresponding to $V$ and we set $\aff(a_i,a_j|Q)=\aff(a_j,a_i|Q)=w_l/\max\{w|e=(u,v,w)\in E\}$ for each edge $e_l=(v_i,v_j,w_l)$. Note that the $\aff$ matrix we set must be legal since the assumed query workload $Q$ which contains $\aff(a_i,a_j|Q)*\max\{w|e=(u,v,w)\in E\}$ queries with predicates only on column $a_i$ and $a_j$ for each $\aff(a_i,a_j|Q)$ value is easy to generate. Then we solve \qpnode construction problem to partition $A$ into $K$ subsets: $\Aset=\{A_1,\ldots,A_K\}$ and we get the answer of the K-CUT problem: minimized $IPA(\Aset|Q)$ multiplying $\max\{w|e=(u,v,w)\in E\}$.

So far, there is no efficient deterministic algorithm for K-CUT problem. When $K=2$, it can be solved by an $O(|A|^3)$ but actually slow MIN-CUT algorithm~\cite{2cut}. When $K=3$, it can be solved by an $O(|A|^3\tau(|A|))$ 3-CUT algorithm~\cite{3cut} where $\tau(|A|)$ is the cost of computing goal function which is $O(|A|^2)$ under the background of our \qpnode construction problem \ie the algorithm costs $O(|A|^5)$. When $K=4$, it can be solved by an $O(|A|^6\tau(|A|))$ \ie $O(|A|^8)$ 4-CUT algorithm~\cite{4cut}. When $K\geq5$, there is no polynomial time algorithm found so far while the research of 4-CUT algorithm~\cite{4cut} made a corollary that the 5-CUT algorithm should cost $O(|A|^{14}\tau(|A|))$ \ie $O(|A|^{16})$. 
\end{proof}	

\subsection{PROOF of LEMMA 2}

\begin{lemma}
	For any given query set \( Q \), the upper bound \( \overline{\ipa}(\mathcal{A} | Q) \) provides an upper estimate of the optimal result \( \ipa(\mathcal{A}^{*} | Q) \).
\end{lemma}

\begin{proof}
	Due to the space constraints, we provide only a proof sketch. First, we show that $\ipa(\mathcal{A}^{*} | Q)$ can be decomposed into the IPA scores of distinct query patterns, \ie
	$
	\ipa(\mathcal{A}^{*} | Q) = \sum_{p_i \in Q}{\ipa(\mathcal{A}^{*} | \{p_i\})}.
	$
	Next, we prove that for any query pattern \( p_i \in Q \), its optimal IPA score \( \ipa(\mathcal{A}^{*} | \{p_i\}) \) is bounded by the upper term \( n_i \cdot z_{ij} \cdot  (\| p_i \| - z_{ij}) \), where \( p_j \) is any query pattern satisfying \( z_{ij} > 0 \). 
	To see why this holds, note that \( z_{ij} \) represents the number of shared accessed columns between patterns \( p_i \) and \( p_j \). The term \( (\| p_i \| - z_{ij}) \) captures the remaining accessed columns in \( p_i \) that are not shared with \( p_j \). Thus, \( n_i \cdot z_{ij} \cdot  (\| p_i \| - z_{ij}) \) is the IPA score of the specific partitioning scheme as illustrated in Figure~\ref{fig:SingleTab_QS_a}, which is larger than the optimal result $\ipa(\mathcal{A}^{*} | \{p_i\})$. 
	Moreover, if \( z_{ij} = 0 \) for all \( j \neq i \), meaning that \( p_i \) shares no common accessed columns with any other query patterns, then the optimal column partition for \( p_i \) incurs no IPA cost, implying \( \ipa(\mathcal{A}^{*} | \{p_i\}) = 0 \). 
	Thus, we conclude that \( \ipa(\mathcal{A}^{*} | Q) \leq \overline{\ipa}(\mathcal{A} | Q) \), proving the lemma.
\end{proof}

\subsection{PROOF of LEMMA 3}

\begin{lemma}
The problem of partitioning \( Q \) into \( \mathcal{Q} = \{Q_1, Q_2, \ldots, Q_n\} \) to minimize the upper bound of inter-partition affinity, \ie, \( \sum_{k=1}^{n} {\overline{\ipa}(\mathcal{A}_{k} | Q_k)} \), is NP-hard.
\end{lemma}

\begin{proof}
	To convert an arbitrary graph $G(V,E)$ to \qsnode problem, we generate $\mathbf{q_i}$ corresponding to each $v_i\in V$ and we generate $w=|\mathbf{q_i}|z_{i,j}(x_i-z_{i,j})+|\mathbf{q_j}|z_{i,j}(x_j-z_{i,j})$ for each $e_{ij}\in E$. To maximize the sum of edge weights that are cut off in $G$, the sum of edge weights remained in all components is minimized, which is equivalent to minimizing $\siq(\mathcal{Q}|A)$ after converting $G$ to \qpnode problem. Thus, minimizing $\siq(\mathcal{Q}|A)$ can be reduced to MAX-CUT problem.
\end{proof}

\subsection{Complexity Analysis of Offline \sys Construction}

Given data $T$, workload $Q$ and column set $|\mathcal{A}|=m$, supposing a balanced $c$-way \sys tree with $n$ nodes ($s$ \sumnode nodes, $x$ \qsnode nodes, $u$ \prodnode | \qpnode nodes, less than and around $n$ \leafnode nodes) and $z$ sample size for RDC. For each \leafnode node, the number of tuples is around $|T|/s$ so its complexity is $O(n|T|/s)$. For each of the data tuples, it will go through $log_cs$ \sumnode nodes and cost $O(cm)$ in each node so the complexity of \sumnode nodes is $O(cm|T|log_cs)$. Similarly, the complexity of \qsnode nodes is $O(cm|Q|log_ct)$. The cases are different with \prodnode and \qpnode, since their cost contains RDC and \aff matrixes computation. For RDC matrix computation, the cost is constant $O(zlogz)$. So the complexity of \prodnode nodes is $O(nm^2zlogz)$. For \aff matrix calculation, the cost is $O(m^2|Q|)$. So the complexity of \qpnode nodes is $O((s+x+u)m^2|Q|)$. In total, the complexity of \sys construction is $O(m^2[nzlogz+(s+x+u)|Q|]+cm(|T|log_cs+|Q|log_cx)+n|T|/s)$. 

\subsection{Complexity Analysis of Online \cardest Inference with \sys}

Given a query \( q \) with selectivity (probability) \( p \) and constraints on \( m \) columns, suppose the \sys model is a balanced \( c \)-way tree with \( N \) nodes (\( s \) \sumnode nodes, \( x \) \qsnode nodes, and \( u \) \prodnode/\qpnode nodes). It is known that the cost of a \leafnode node with a histogram is \( O(1) \), the cost of a \qsnode node is \( O(m^2 c) \), and the cost of other middle nodes is \( O(m c) \). Thus, the worst-case complexity of the bottom-up inference method is \( O(x m^2 c + (N - x) m c) \). Since there are usually a small number of \qsnode nodes in a \sys model for query routing, the time cost of our fast inference method is approximately \( \max(p, \frac{1}{s}) \cdot \frac{m}{u} \cdot N m c \).

%% file: secs/appendixC.tex
\section{Additional Detailed Algorithm}

\subsection{Offline \sys Construction Algorithm}

\input{algs/cardest-single-offline-simple}

We show our top-down \att{Construct\sys} (Algorithm 1) corresponding to \Cref{sec:qspn-offline} including its sub function \att{PartitionByAFF} (Algorithm 1.1) which is explained in \Cref{subsec:construct-qproduct} and sub function \att{SplitWorkload} (Algorithm 1.2) which is explained in \Cref{subsec:construct-qsplit}. In \att{Construct\sys} algorithm, each node $\node$ to construct are tried different type $\node.O$ in the heuristic order: \leafnode, \prodnode, \qpnode, \qsnode, \sumnode for a accuracy-efficiency balanced \sys model. And the specific node construction are executed in sub functions.

\subsection{Online \sys Inference Algorithm}

\input{algs/cardest-single-online-simple}

We show our top-down \att{QSPN-Online} (Algorithm 2) based on query routing and pruning rules whose sub function \att{RouteQuery} is explained in \Cref{subsec:qspn-online}. Benefiting from sufficient optimization utilizing the features of \sys, the \att{QSPN-Online} algorithm provides accurate and fast \cardest in online phase using lightweight \sys model which are constructed both on data and workload before.

\subsection{Online \sys Update Algorithm}

\input{algs/cardest-single-update}

We show our efficient top-down \att{QSPN-Update} (Algorithm 3) based on node examining whose sub function \att{ColumnPartitionCheck}, \att{WorkloadSplitCheck} and \att{DataPartitionCheck} are explained in \Cref{subsec:qspn-update}. The \att{QSPN-Update} algorithm achieves adaptive incremental \sys update to keep the model accurate against data update and query workload shift in online phase of dynamic scenes, which is manual-triggering-free and brings little extra cost.

\subsection{\mqspn Multi-Table \cardest Algorithm}

\input{algs/cardest-multi-online-simple}

We show our \att{\mqspn-BinningGen} (Algorithm 4.1) and \att{\mqspn-Multi\cardest} (Algorithm 4.2) based on our \mqspn model which are explained in \Cref{sec:qspn-multi}. Benefiting from outstanding single-table \sys model and join-key binning framework, these two algorithms achieve accurate and efficient multi-table \cardest together with hightlights: no outer-join table generation and no model training on outer-join table.

%% file: algs/cardest-single-offline-simple.tex
\renewcommand{\thealgorithm}{1} 
\begin{algorithm}
	\caption{\att{Construct}\sys $(\Aset, T, Q)$}
	\begin{algorithmic}[1] 
		\Require $\Aset$: A Column Set; $T$: A Table; $Q$: A Query Workload
		\Ensure $\node$: A \sys Node 
		\State Initialize a \sys node $\node=(A, T, Q, O)$ 
		\If{$|\Aset|=1$} 
			\State $\node.O \gets \texttt{Leaf}$
			\State $\node.Pr_{T}(\Aset)$ $\gets$ \att{BuildHist} ($A$, $T$)
		\ElsIf{$\mathcal{A} \gets$ \att{PartitionByRDC} ($A$, $T$) is \emph{non-singleton}}
			\State $\node.O \gets \texttt{Product}$
			\State $\node.\texttt{child}_i \gets$ \att{Construct}\sys $(A_i, T, Q)$ for $\forall A_i \in \mathcal{A}$
		\ElsIf{$\mathcal{A} \gets$ \att{PartitionByAFF} ($A$, $Q$) is \emph{non-singleton}}
			\State $\node.O \gets \texttt{Product}$
			\State $\node.\texttt{child}_i \gets$ \att{Construct}\sys $(A_i, T, Q)$ for $\forall A_i \in \mathcal{A}$
		\ElsIf{$\mathcal{Q} \gets$ \att{SplitWorkload} $(Q, A)$ is \emph{non-singleton}}
			\State $\node.O \gets \texttt{QSplit}$
			\State $\node.\texttt{child}_i \gets$  \att{Construct}\sys $(A, T, Q_i)$ for each $Q_i \in \mathcal{Q}$
		\Else 
            \State $\node.O \gets \texttt{Sum}$
			\State $\mathcal{T} \gets$ \att{ClusterTable} $(T, A)$
			\State Compute weight $w_i \gets {|T_i|} / {|T|}$ for each sub-table $T_i \in \mathcal{T}$
			\State $\node.\texttt{child}_i \gets$  \att{Construct}\sys $(A, T_i, Q)$ for each $T_i \in \mathcal{T}$
		\EndIf
		\State \Return node $\node$
	\end{algorithmic}
\end{algorithm}

\renewcommand{\thealgorithm}{1.1} 
\begin{algorithm}
	\caption{\att{PartitionByAFF} $(Q, A)$} 
	\begin{algorithmic}[1] 
		\Require $Q$: A Query Workload; $A$: A Column Set
		\Ensure $\mathcal{A} = \{A_1, \ldots, A_m\}$: A Collection of Column Sets
		\For{any pair $(a_i, a_j)$ from $A$}
			\State Compute $\aff(a_i, a_j | Q)$ for $a_i$ and $a_j$
		\EndFor 
		\State Construct a graph $G = (A, E)$ where vertices $A$ represent columns, and an edge $e_{ij} \in E$ exists between columns $a_i$ and $a_j$ if their affinity $\aff(a_i, a_j | Q)$ is larger than a threshold $\tau$.
		\State $\mathcal{G} \gets $\att{ConnectedComponents} $(G)$
		\State \Return $\mathcal{A}=\{A_i \mid$ vertex set in each component $G_i \in \mathcal{G}$$\}$
	\end{algorithmic}
\end{algorithm}

\renewcommand{\thealgorithm}{1.2} 
\begin{algorithm}
	\caption{\att{SplitWorkload} $(Q, A)$} 
	\begin{algorithmic}[1] 
		\Require $Q$: A Query Workload; $A$: A Column Set 
		\Ensure $\mathcal{Q}=\{Q_1, \ldots Q_m\}$: A Collection of Sub-workloads
        \State $\mathbf{q_1},\ldots,\mathbf{q_w} \gets$ Merge Queries $q$ sharing the same $\accvec_{q}$
        \State $\simi\leftarrow$ \att{ComputeSim} ($\ACC(Q, \Aset)$) for each $(q_i,q_j)$ from $Q$
        \State Construct a graph $G=(V,E)$ where vertice $V$ represents $\ACC(Q,\Aset)_1,\ldots,\ACC(Q,\Aset)_{|\ACC(Q,\Aset)|}$ and $e_{kl}\in E$ represents $\simi(\ACC(Q,A)_k,\ACC(Q,A)_l)$
        \State Cut off some biggests edges of $G$
        \State Sort $v\in V$ by weighted degrees in desc order
        \State Create $m$ empty set: $R_1,\ldots,R_m$
        \For{$1\leq i\leq |\ACC(Q,\Aset)|$}
                \State Put $\ACC(Q,\Aset)_i$ into the subset $R_j$ with $\min_j(\sum_{\ACC(Q,A)_i\in R_j}\simi(i,\ACC(Q,A)_k))$
        \EndFor
        \State Create $m$ empty query subsets: $Q_1,\ldots,Q_m$
        \For{$q\in Q$}
            \State Put $q$ into $Q_k$ if $\exists\ACC(Q,\Aset)_l\in R_k$ and $q$ obeys $\ACC(Q,\Aset)_l$
        \EndFor
        \State $\mathcal{Q}\leftarrow\{Q_1,\ldots,Q_m\}$
        \State \Return $\mathcal{Q}$

	\end{algorithmic}
\end{algorithm}

%
%

%% file: algs/cardest-single-online-simple.tex
\renewcommand{\thealgorithm}{2} 
\begin{algorithm}[t!]
	\caption{\att{QSPN-Online} $(\node, q)$}
	\begin{algorithmic}[1] 
		\Require $\node$: A QSPN Node; $q$: A Query
		\Ensure $P_\node$: Probability Estimated at Node $\node$  
		\If{$\node.O = \texttt{Leaf}$}
			\State \Return $\node.P_T(q)$ based on the histogram at leaf $\node$
		\ElsIf{$\node.O = \texttt{Product} \mid \texttt{QProduct}$}
			\State $P_i \gets$  \att{QSPN-Online} $(\node.\texttt{child}_i, q)$ for each child of $\node$ with the columns satisfying $q.A \cap \node.\texttt{child}_i.A \neq \emptyset$
			\State \Return $\prod_{i}{P_i}$ from all child nodes of $\node$
		\ElsIf{$\node.O = \texttt{QSplit}$}
			\State $k \gets \att{RouteQuery}~(q, \{\node.\texttt{child}_i\})$
			\State \Return  \att{QSPN-Online} $(\node.\texttt{child}_k, q)$
		\Else
			\State $\mathcal{C} \gets $ \att{SelectChildren} ($\{\node.\texttt{child}_i\}, q$)
            \State $P_i \gets$  \att{QSPN-Online} $(\texttt{child}, q)$ for each $\texttt{child} \in \mathcal{C}$
        	\State \Return weighted summation $\sum_{i}{w_i \cdot P_i}$
		\EndIf
	\end{algorithmic}
\end{algorithm}

%% file: algs/cardest-single-update.tex
\renewcommand{\thealgorithm}{3} 
\begin{algorithm}[t!]
	\caption{\att{QSPN-Update} $(\node,\Delta T,\Delta Q)$} 
	\begin{algorithmic}[1] 
		\Require $\node$: A \sys Node, $\Delta T$: Data Update subset, $\Delta Q$: New Queries subset
		\Ensure Updated QSPN node $\node'$
            \State $U\gets False$
            \If{$\node.O=\leafnode$}
                \State $\node'\gets\node$
                \State Modify Histogram at $\node'$ by $\Delta T$
                \State \Return $\node'$
            \ElsIf{$\node.O=\prodnode\mid\qpnode$}
                \State $U,\node'\gets$\att{ColumnPartitionCheck} $(\node,\Delta T,\Delta Q)$
            \ElsIf{$\node.O=\qsnode$}
                \State $U,\node'\gets$\att{WorkloadSplitCheck} $(\node,\Delta Q)$
            \ElsIf{$\node.O=\sumnode$}
                \State $U,\node'\gets$\att{DataPartitionCheck} $(\node,\Delta T)$
            \EndIf
            \If{$U$}
            \State $\node'\gets \att{QSPN-Offline}$ $(\node'.\Aset,\node'.T\cup\Delta T,\node'.Q\cup\Delta Q)$
            \Else
            \State $\node'.\child_i\gets \att{QSPN-Update}$ $(\node'.\child_i,\Delta T_i,\Delta Q_i)$ for each child of $\node'$
            \EndIf
		\State \Return $\node'$
	\end{algorithmic}
\end{algorithm}

%% file: algs/cardest-multi-online-simple.tex
\renewcommand{\thealgorithm}{4.1} 
\begin{algorithm}
	\caption{\att{\mqspn-BinningGen}} 
	\begin{algorithmic}[1] 
		\Require $\node$: A M-QSPN Node on Table $T$; $q$: A Query; $id$: Join Key
		\Ensure $B^T$: Table $T$ Binning, $P_\node$: Probability Estimated at Node $\node$ 
            \State Compute $\node.P_T(q)$ in the same way of \att{QSPN-Online} $(\node, q)$
            \If{$id\in\node.A$}
    		\If{$\node.O = \texttt{Leaf}$}
                    \State Generate $q$-filtered $B^T$ based on the histogram at leaf $\node$
                \Else
                    \State Generate $B^T$ by merging $B^T_i$ by \att{\mqspn-BinningGen} $(\node.\texttt{child}_i,q,id)$ following strategies of different $\node.O$
                \EndIf
            \Else
                \State $B^T\gets\emptyset$
            \EndIf
            \State \Return $B^T,\node.P_T(q)$
	\end{algorithmic}
\end{algorithm}

\renewcommand{\thealgorithm}{4.2} 
\begin{algorithm}
	\caption{\att{\mqspn-Multi\cardest}} 
	\begin{algorithmic}[1] 
		\Require $B^S,B^T$: Binnings corresponding to table $S$ and $T$
		\Ensure Multi-table Cardinality $card$
            \For{bin $b\in B^S$}
                \State Find $b'\in B^T$ sharing same range of $b$
                \If{$b \text{ shares same }\texttt{mcv}\text{ with } b'$}
                    \State $card_b\gets b.\texttt{mcv}\times b'.\texttt{mcv}$
                    \If{$b.\texttt{mcv} << b.\texttt{num} \land b'.\texttt{mcv} << b'.\texttt{num}$}
                        \State $card_b\gets card_b\times\min(\frac{b.\texttt{num}}{b.\texttt{mcv}},\frac{b'.\texttt{num}}{b'.\texttt{mcv}})$
                    \EndIf
                \Else
                    \If{$b.\texttt{mcv}>b'.\texttt{mcv}$}
                        \State $card_b\gets b.\texttt{mcv}\times\frac{b'.\texttt{num}-b'.\texttt{mcv}}{|b'.\texttt{range}|-1}$
                    \Else
                        \State $card_b\gets \frac{b.\texttt{num}-b.\texttt{mcv}}{|b.\texttt{range}|-1}\times b'.\texttt{mcv}$
                    \EndIf
                \EndIf
                \State $card\gets card+card_b$
            \EndFor
            \State \Return $card$
	\end{algorithmic}
\end{algorithm}